\documentclass[referee]{sn-jnl}

\usepackage[version=3]{mhchem} 
\usepackage{amsmath,amssymb,amsfonts}
\usepackage{xcolor}
\usepackage{graphicx}
\usepackage{tabularx}
\usepackage{listings}%
\usepackage{cite}

\usepackage{xr}
\makeatletter
\newcommand*{\addFileDependency}[1]{%
    \typeout{(#1)}
    \@addtofilelist{#1}
    \IfFileExists{#1}{}{\typeout{No file #1}}
}
\makeatother

\newcommand*{\myexternaldocument}[1]{%
    \externaldocument{#1}%
    \addFileDependency{#1.tex}%
    \addFileDependency{#1.aux}%
}
\myexternaldocument{supplementary}

\usepackage{siunitx} 
\usepackage{cleveref}

\begin{document}

\title{Peering inside the black box: Learning the relevance of many-body functions in Neural Network potentials}

\author[1]{\fnm{Klara} \sur{Bonneau}}
\equalcont{These authors contributed equally to this work.}

\author[2,3]{\fnm{Jonas} \sur{Lederer}}
\equalcont{These authors contributed equally to this work.}

\author*[1]{\fnm{Clark} \sur{Templeton}}\email{clarktemple03@gmail.com}
\equalcont{These authors contributed equally to this work.}

\author[1]{\fnm{David} \sur{Rosenberger}}

\author*[2,3,4,5]{\fnm{Klaus-Robert} \sur{M\"uller}}\email{klaus-robert.mueller@tu-berlin.de}

\author*[1,6,7]{\fnm{Cecilia} \sur{Clementi}}\email{cecilia.clementi@fu-berlin.de}

\affil[1]{\orgdiv{Department of Physics}, \orgname{Freie Universit\"at Berlin}, \orgaddress{\street{Arnimallee 12}, \postcode{14195}, \city{Berlin}, \country{Germany}}}

\affil[2]{\orgdiv{Machine Learning Group}, \orgname{Technische Universit\"at Berlin}, \orgaddress{\street{Marchstr. 23}, \postcode{10587}, \city{Berlin},  \country{Germany}}}

\affil[3]{\orgdiv{BIFOLD - Berlin Institute for the Foundations of Learning and Data}, \orgaddress{\country{Germany}}}

\affil[4]{\orgdiv{Department of Artificial Intelligence, Korea University}, \orgname{Korea University}, \orgaddress{\city{Seoul}, \street{136-713},  \country{South-Korea}}}

\affil[5]{\orgname{Max Planck Institute for Informatics}, \orgaddress{\postcode{66123}, \city{Saarbr\"ucken}, \country{Germany}}}

\affil[6]{\orgdiv{Center for Theoretical Biological Physics}, \orgname{Rice University}, \orgaddress{\city{Houston}, \postcode{77005}, \state{TX}, \country{USA}}}

\affil[7]{\orgdiv{Department of Chemistry}, \orgname{Rice University}, \orgaddress{\city{Houston}, \postcode{77005}, \state{TX}, \country{USA}}}

\abstract{
Machine learned potentials are becoming a popular tool to define an effective energy model for complex systems, either incorporating electronic structure effects at the atomistic resolution, or effectively renormalizing part of the atomistic degrees of freedom at a coarse-grained resolution.
One of the  main criticisms to machine learned potentials is that the energy inferred by the network is not as interpretable as in more traditional approaches where a simpler functional form is used.
Here we address this problem by extending tools recently proposed in the nascent field of Explainable Artificial Intelligence (XAI) to coarse-grained potentials based on graph neural networks (GNN). 
We demonstrate the approach on three different coarse-grained systems including two fluids (methane and water) and the protein NTL9. On these examples, we show that the neural network potentials can be in practice decomposed in relevance contributions to different orders, that can be directly interpreted and provide physical insights on the systems of interest.}


\maketitle

\section{Introduction}\label{sec:intro}
Molecular simulations have emerged in the last 75 years as a valuable tool to recover or even predict interesting physical phenomena at the microscopic scale and provide a detailed mechanism for grasping the underlying molecular processes~\cite{best_atomistic_2019}.
In principle, the most accurate description of a molecular system is given by the solution of the associated Schrödinger's equation. 
However, it is common practice to invoke the separation of scales between electrons and nuclei (Born-Oppenheimer approximation) and define an effective energy function for the nuclei that should take into account the electronic effects~\cite{monticelli2013force}. 
Historically, this has been done empirically in the definition of classical atomistic force-fields, which have been designed, refined, and used for the study of molecular systems~\cite{lindorff-larsen_systematic_2012,robustelli_developing_2018}.   
Classical force fields assume that the energy of a molecular system can be described as a function of ``bonded" terms (e.g. bonds, angles, dihedrals) and ``non-bonded" pairwise potentials (e.g., Van der Waals, Coulomb)~\cite{monticelli2013force,best_atomistic_2019}. 
All the potential energy terms are defined by fixed functional forms, with parameters tuned to reproduce experimental data and/or first principle calculations on small test systems~\cite{lindorff-larsen_systematic_2012,best_atomistic_2019}.

Recent advances in machine learning have triggered a step-change in the development of data-driven force fields. In particular, Artificial Neural Networks (ANNs) have been proposed to more accurately capture the electronic effects in the potential energy functions for the nuclei~\cite{behler_generalized_2007,noe_machine_2020}. 
While classical, non-bonded potential terms are generally limited to 2-body interactions, the use of ANNs, and more specifically Graph Neural Networks (GNNs)~\cite{scarselli_graph_2009} defining connections between neighboring atoms, significantly increases the expressivity of the energy function and allows a flexible parameterization of many-body interactions~\cite{duvenaud_convolutional_2015,kearnes_molecular_2016,gilmer2017neural,schuett2017dtnn,schutt_schnet_2018,unke2019physnet}.

While leaps in the development of GNN-based models have shown great promise in studying complex macromolecular systems \cite{Unke2024} and predicting material properties~\cite{ceriotti2022beyond}, the results and the models themselves are often seen as black boxes.
In a machine learned force-field, a molecular conformation is given as input to the neural network and only the total energy and its derivatives are obtained as output.
The increased model accuracy comes at the cost of insight into the nature and strength of molecular interactions:
In a classical force field each term in the energy function can be dissected, but deciphering which terms in the potential energy are important for stabilizing certain physical states or interpreting a prediction is significantly more difficult in a GNN-based model.

This black box problem is not unique to molecular systems, but rather ubiquitous in the application of machine learning.
As a response, the new area of ``Explainable Artificial Intelligence (XAI)" has emerged to start providing tools to tackle the interpretation of deep neural networks \cite{samek21xaireview}. 
Different approaches have been proposed to interpret the results obtained with ANNs, ranging from self-explainable architectures~\cite{schuett2019xaiqc, schuett2017dtnn, gallegos_explainable_2024, zhang2022protgnn} 
to post-hoc explanations
~\cite{lundberg17unified, bach2015pixel, sundararajan2017axiomatic, ribeiro2016should}. 
Some of those approaches are starting to find use also in physical and chemical applications, 
e.g., for explaining predictions regarding toxicity or mutagenicity~\cite{debnath1991structure, Kazius2005DerivMutag, baehrens2010explain}, 
predictions of electronic-structure properties~\cite{letzgus2022toward, schnake2022gnnlrp}, guiding strategies in drug discovery~\cite{jose20drugdisc, wellawatte2023perspective}, analyzing protein-ligand binding~\cite{mcclo19usingatt, lill2019elucidating, hochuli2018visualizing}, 
or uncertainty attribution~\cite{yang2023explainable, roy2023learning}. XAI approaches have also been utilized to provide a better understanding of the error introduced in coarse-grained molecular models~\cite{durumeric2023using}.

In principle, an ``interpretable model" should allow a researcher to extract scientific knowledge in a successful application and to identify the sources of deficiencies/anomalies when the model fails. In this work, we use ideas from  XAI and implement them in the context of machine learned molecular models for molecular dynamics simulations. 
In particular, we focus our attention on the understanding of coarse-grained (CG) models. In parallel to the development of atomistic force-fields, GNNs have been successfully employed in the definition of models at reduced resolutions \cite{jiang2017coarse,zhang_deepcg_2018,husic2020coarse,majewski2023machine,charron2023}, where some of the atomistic degrees of freedom are renormalized into a reduced number of effective ``beads" to speed up simulation time.
The difficulty in the definition of CG models lies in the fact that many-body terms play an important role, as a reduction in the number of degrees of freedom is associated with increased complexity in the effective CG energy function.
It has been shown that to reproduce either experimentally measured free energy differences \cite{zaporozhets2023multibody} or the thermodynamics of a finer-grained model \cite{larini_coarse-grained_2012,wang2021multi}, many-body terms need to be included. 
The need for many-body terms and the difficulty in capturing them make CG models an ideal test ground for GNN interpretation.

In the present study, we train a GNN energy function at CG resolution from atomistic simulation data of several systems of different complexity and then interpret the model to provide a deeper understanding, rather than a mere energy prediction. 
By using explainable AI in the context of CG models, we provide evidence that machine learned force-fields are indeed learning physically significant interactions. 

To interpret the energy prediction of a CG molecular model, we use the method of ``Layerwise-relevance propagation" (LRP) for GNNs, which has been recently proposed to explain a model prediction by decomposing the output (in our case, the energy of the system) in terms of the contribution (or ``relevance") of groups of nodes (i.e. CG beads)~\cite{schnake2022gnnlrp}. 
In a sense, we can see the GNN-LRP as a multi-body expansion of the total energy of the system provided by the GNN.

As a first example, we compare different classes of GNN architectures to obtain CG models of bulk fluids and show that an interpretation of accurate CG models provide a meaningful physical information. Interestingly, two GNN architectures, even if different from each other, convey the same physical interpretation: at least in terms of 2-body and 3-body terms, the two networks offer different functional representations of the same underlying energy landscape. 

As a second example, we examine a machine learned CG model for the protein NTL9 and show that its interpretation allows us to pinpoint the stabilizing and destabilizing interactions in the various metastable states, and even interpret the effects of mutations. 

\section*{Results}
\subsection{Methane and Water}
We start by analyzing and comparing CG models for bulk methane (CH$_4$) and water (H$_2$O). 
Methane has been previously studied with various coarse-graining methods since its non-polar and weak Van-der-Waals interactions make it a simple test system~\cite{jiang2017coarse}.
On the other hand, water is capable of forming complex hydrogen bonding structures and much research has been devoted to its modelling, both at the atomistic scale~\cite{ouyang_modelling_2015} and on the CG level~\cite{molinero_water_2009, noid2008mscg1,larini2010multiscale,das_multiscale_2012, zhang_deepcg_2018}.

For both systems, we define CG models by integrating out the hydrogen atoms and positioning an effective CG bead in place of the central carbon or oxygen atom.
For each system, we train two CG models with different choices of GNN architectures, PaiNN~\cite{schutt2021equivariant} and SO3Net~\cite{schutt2023schnetpack}, for the definition of the CG effective energy.
Both architectures are trained using the force-matching variational principle for coarse-graining, to create a thermodynamically consistent CG model from the atomistic data~\cite{ercolessi_interatomic_1994, izvekov_multiscale_2005, noid2008mscg1, wang_machine_2019, husic2020coarse}.
A brief discussion of the specific features of these GNNs is provided in the Methods section.

\begin{figure}[h]
\centering
\includegraphics[width=\linewidth]{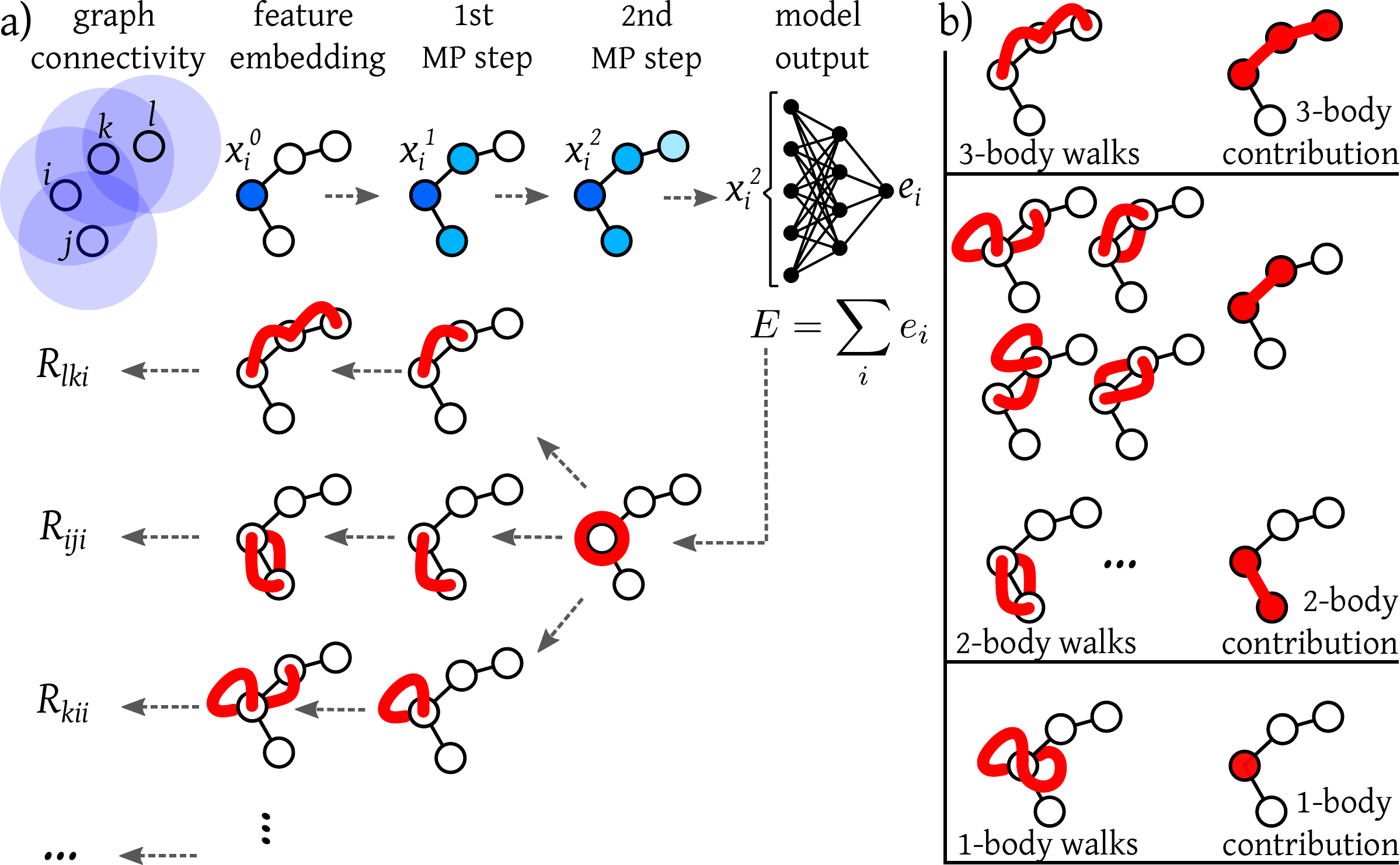}
\caption{Concept of GNN-LRP illustrated for a system of four particles (i.e. CG beads, in the present context). a) 
In GNNs, the input graph is defined by a cutoff radius that determines the direct neighbors for each input node. By stacking several message aggregations in multiple layers, information can be exchanged between more distant nodes (outside of the cutoff region). 
The model output is obtained by passing the learned feature representations through a multilayer perceptron and final pooling. Obtaining the relevance involves propagating the output back through the network, by considering the connections between each node in one layer and the nodes in the previous layer. This procedure defines ``walks" across the network layers. b) The walks involving the same subset of $n$ nodes are aggregated to obtain a decomposition of the output into $n$-body contributions.
}
\label{fig:LRPUnderstanding}
\end{figure}

To interpret the network predictions, we use the method of  ``Layerwise-relevance propagation" (LRP) for GNN, called GNN-LRP~\cite{schnake2022gnnlrp}. 
The idea of GNN-LRP is schematically described in Fig.~\ref{fig:LRPUnderstanding} and
a more detailed description is provided in the Methods section as well as in the Supplementary Section S5.
Essentially, GNN-LRP provides us with a ``relevance score" for each subset of $n$ CG beads (with $n=1,\dots,N_{l}+1$, where $N_{l}$ is the number of layers in the GNN) in each configuration of the system, indicating the contribution of their interaction to the total energy.  

\begin{figure}[h]
\centering
\includegraphics[width=1.0\linewidth]{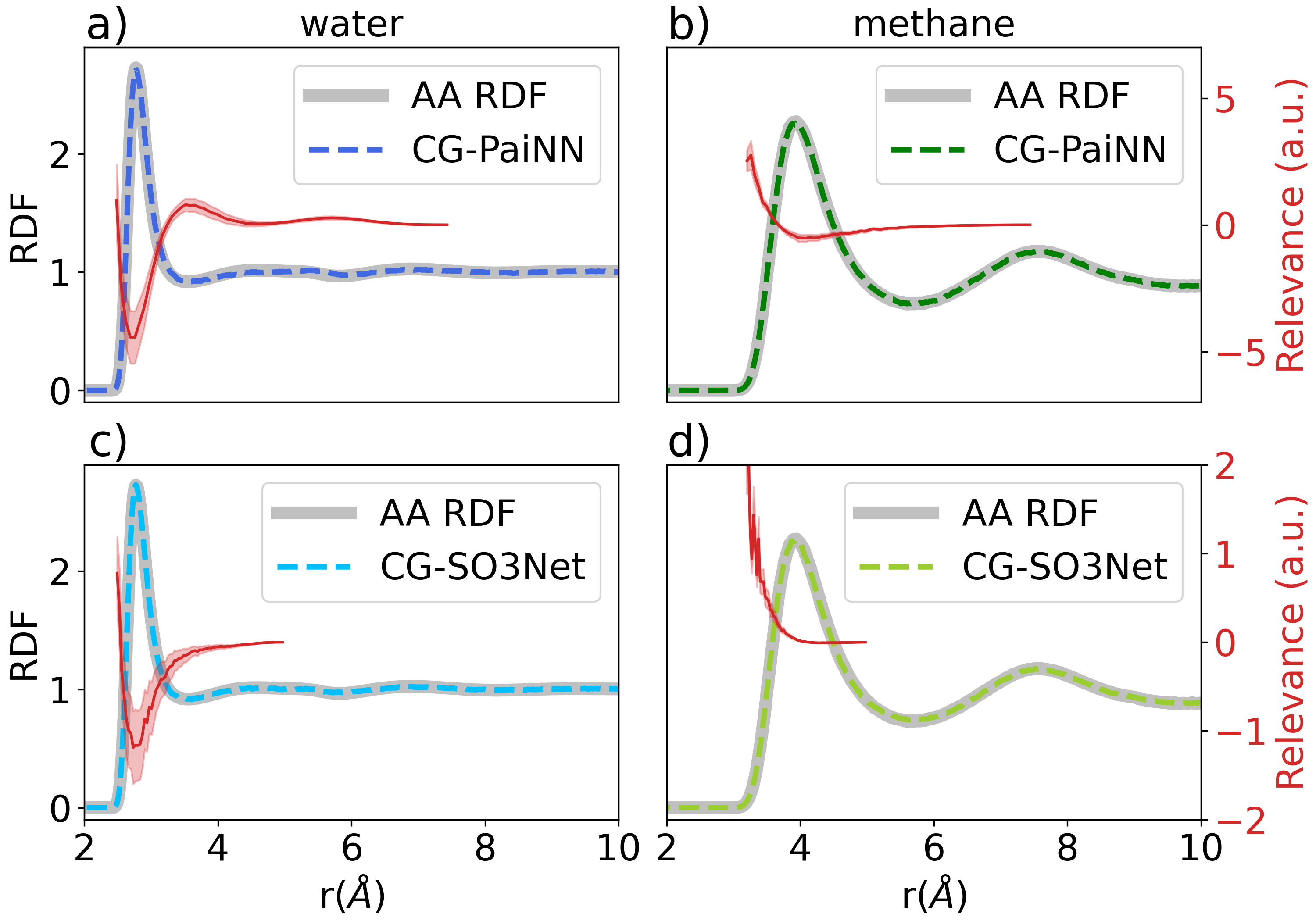}
\caption{Comparison of radial distribution functions resulting from simulations with an atomistic or CG model and corresponding 2-body relevance. Panels a) and c) correspond to water and b) and d) to methane models. Panels a) and b) show the results for PaiNN-based, and panels c) and d) for SO3Net-based models. The relevance, shown in red, is normalized by the absolute total relevance over the number of walks of the respective model and rescaled for each model type. A negative value implies a stabilizing interaction as the model output is the energy.}
\label{fig:MethaneWaterRDF}
\end{figure}

\begin{figure}[t]
\centering
\includegraphics[width=1.0\linewidth]{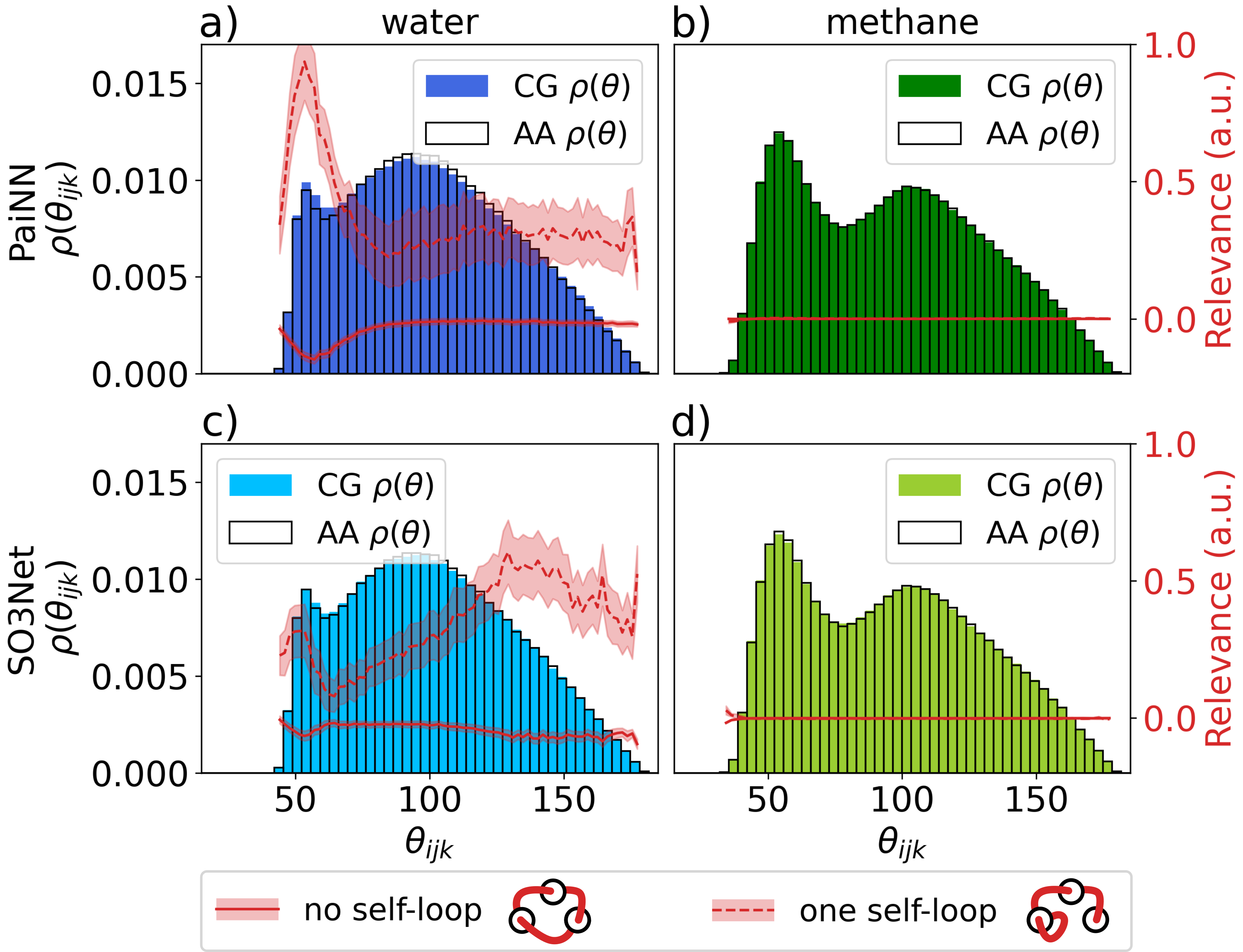}
\caption{3-body relevance for water and methane with similar layout to Fig.~\ref{fig:MethaneWaterRDF} but examining the angular distribution between triplets of neighboring atoms. The relevance is separated into walks containing a self-loop (dashed red line) and walks without self-loops (full red line). 
Angle distributions are computed with a cutoff after the first solvation shell, at 3.5$\mathrm{\AA}$ for water and 5.6$\mathrm{\AA}$ for methane. Only relevance values for triplets within this cutoff are plotted, more details can be found in Supplementary Section S3.}
\label{fig:MethaneWaterAngle}
\end{figure}

The ability of the two CG models to reproduce structural features of the two systems is shown in Figs.~\ref{fig:MethaneWaterRDF}-\ref{fig:MethaneWaterAngle}. 
In particular, the radial distribution function (RDF) as obtained in the CG models is shown in Fig.~\ref{fig:MethaneWaterRDF} against the atomistic reference model for methane (right column) and water (left column). Note that the SO3Net model exhibits irreducible representations up to an angular momentum of $l_\mathrm{max}=2$, while PaiNN utilizes a maximum angular momentum of $l_\mathrm{max}=1$. As a consequence, in comparison to SO3Net, PaiNN requires a larger cutoff to accurately reproduce the RDF of water. For a comparison between PaiNN models with different cutoff radii, please refer to Supplementary Fig.~S2.

For methane, the smooth oscillatory behavior of the RDF is similar to a Lennard-Jones (LJ) fluid ~\cite{mcquarrie76a}, suggesting that many-body interactions may not be very relevant in a CG model of this molecule.
In contrast, for water, the height of the first solvation shell is more sharply peaked and decays more rapidly than in the case of methane. 
For both systems, both architectures reproduce the corresponding RDF. 

In Fig. \ref{fig:MethaneWaterRDF}, the average relevance score for the 2-body contributions is plotted (red curves) as a function of their distance, alongside the RDF for each model.
Since the relevance score corresponds essentially to a decomposition of the output energy,
a positive (negative) relevance score implies an increase (decrease) in the energy, thus a destabilizing (stabilizing) effect of the associated interaction. 

For both methane and water, both PaiNN and SO3Net show a 2-body relevance score that diverges as the distance between two beads goes to zero. 
This observation matches our intuition that at distances below a certain ``effective radius", the network should learn a repulsive excluded-volume  interaction to avoid the overlapping of the CG beads. 
For all models, the relevance score decays to zero as the distance between two atoms approaches the cutoff value considering two atoms connected in the corresponding GNN. This corresponds to the intuition that the interactions between two atoms become weaker as the atoms move further apart, and is enforced by the cosine shape of the cutoff function.

Differences between the two systems emerge at intermediate distances.
Fig. \ref{fig:MethaneWaterRDF}b) and d) shows that, for methane, the relevance score is mostly positive. The PaiNN model displays a slight stabilizing well around the first peak of the RDF, that is washed out in the SO3Net model due to the proximity between the network cutoff and the location of the first solvation shell.
On the other hand, the relevance score for both water models (Fig. \ref{fig:MethaneWaterRDF} a) and c)) dips significantly below 0 at $\sim\SI{3}{\angstrom}$, indicating a stabilizing interaction between molecules at a distance corresponding to the first solvation shell. 

Fig. \ref{fig:MethaneWaterAngle} examines the angle distributions between triplets of interacting atoms within the first solvation shell, and reports the average 3-body relevance score as a function of this angle (as red curves). As illustrated in Fig.~\ref{fig:LRPUnderstanding}, the number of steps in the ``walks" of connections between the nodes (CG beads) from one layer to the previous one in the GNN is determined by the number of message passing interaction blocks of the ML model (see Methods section for details). Here, we use models with three interaction blocks in both PaiNN and SO3Net models, thus we obtain walks of four steps in the input graph (with corresponding relevance attribution \(R_{ijkl}\)). 
As the architecture of both models include skip connections in their interaction blocks, a given graph node in a network layer may be connected to the same node in the previous layer. That is, a node may appear more than once in a single network walk. We indicate the case where a node is connected to itself in two subsequent layers as a ``self loop" (see Fig.~\ref{fig:LRPUnderstanding}). Since features involved in self loops are expected to dominate the respective walk relevance, here, we distinguish between 3-body walks containing a self-loop (eg. connecting nodes [i, i, j, k] in Fig.~\ref{fig:LRPUnderstanding}) and walks containing no self-loops (eg. connecting nodes [i, j, k, i] in Fig.~\ref{fig:LRPUnderstanding}). 

For methane, both PaiNN and SO3Net can reproduce the local atomistic angle distribution and they both have an associated relevance score very close to zero that does not particularly favor any angular configuration, indicating that the 3-body terms are not very important for the CG methane. 
For water, both models produce a similar behavior for the averaged 3-body relevance (shown in red). The walks without self-loop are stabilizing and favor angles around ~$50$-$60$ degrees, corresponding to the population of water molecules sitting interstitially inside the tetrahedral arrangement~\cite{stock_unraveling_2017,monroe_decoding_2019}. The walks containing self-loops are mostly destabilizing and likely correct for an overstructuring of the 2-body interactions. Indeed, in Supplementary Fig. S5, one can see that the relevance from the walks with self-loops varies much more as a function of the distance than the angle, whereas the relevance of walks without self-loops depends strongly on the angles and is not merely dominated by pairwise distances. Having 3-body interactions correct for 2-body interactions is a known effect when parametrizing explicit $n$-body functions for constructing CG models~\cite{larini2010multiscale,das_multiscale_2012,scherer2018understanding}. In both models presented here, 3-body terms are crucial to recover structural properties of CG water.

To further support this point, a 2-body-only model is shown in the Supplementary Section S3, where the Inverse Monte Carlo (IMC) method is used for parametrizing a pair potential on the system's RDF. Supplementary Fig. S1 shows that, for methane, the IMC model is capable of reproducing the correct distributions of the relevant features, whereas for water, this 2-body-only model fails at reproducing the angular distribution.
Additionally, we also observe that a non-equivariant GNN such as SchNet performs well on methane, but fails at fully reproducing the atomistic distributions for water (see Supplementary Section S3).

It's worth noting that while the average relevance is shown in Fig. \ref{fig:MethaneWaterRDF} and Fig. \ref{fig:MethaneWaterAngle} as a function of distances and angles, the individual relevance scores cover a broad range. 
Supplementary Figs. S3 and S4 show the entire distribution of 2-body relevance values over all the configurations of the water models. High (destabilizing) relevance values correspond to short distances between beads and low (stabilizing) relevance values to distances corresponding to the first peak of the RDF. A relevance score of almost zero corresponds to distances approaching the network cutoff.

\subsection{NTL9}
\begin{figure}[h]
\centering
\includegraphics[width=1.0\linewidth]{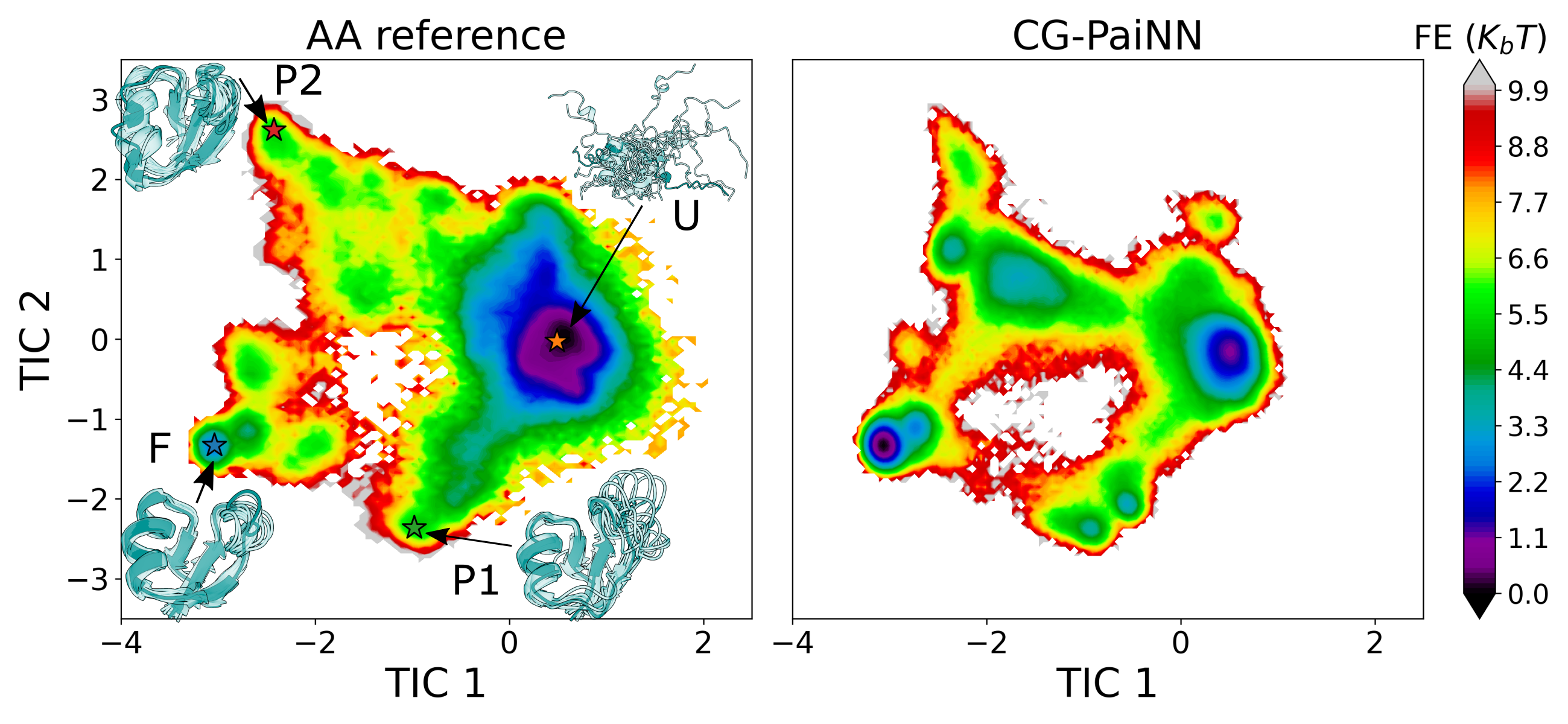}
\caption{Comparison of the FES from the all-atom (left) to the CG (right) shown as a function of the first two TICA components \cite{perez2013identification}. Four regions of interest in TICA space are labeled by F (folded), U (unfolded), P1 (folding pathway 1), and P2 (folding pathway 2). The structures used for the interpretation of the respective states are shown on the left.}
\label{fig:NTL9FES}
\end{figure}

Finally, we show the GNN-LRP interpretation in a coarse-grained protein model, specifically that of 39 residue NTL9 (PDB ID: 2HBA, residues 1-39) which has been a system of interest in a recent study by Majewski~et.~al.~\cite{majewski2023machine}. 
We use the PaiNN architecture introduced in the previous section to learn a CG model of NTL9 from the same atomistic reference data as used in the previous study~\cite{majewski2023machine}, following the procedure introduced by Husic~et.~al.~\cite{husic2020coarse}. More details on the training procedure can be found in the Supplementary Section S2. In this study, only $C_{\alpha}$ atoms are kept in the CG resolution and each amino-acid type is represented by a unique bead embedding. 
We selected this protein due to its well-characterized folded/unfolded state and non-trivial folding pathways.
Comparison of the Free Energy surface (FES) projected onto the first two TICA components (collective variables capturing the slow motions of the system) \cite{perez2013identification} is shown in Fig \ref{fig:NTL9FES}. 

\begin{figure}[h]
\centering
\includegraphics[width=1.0\linewidth]{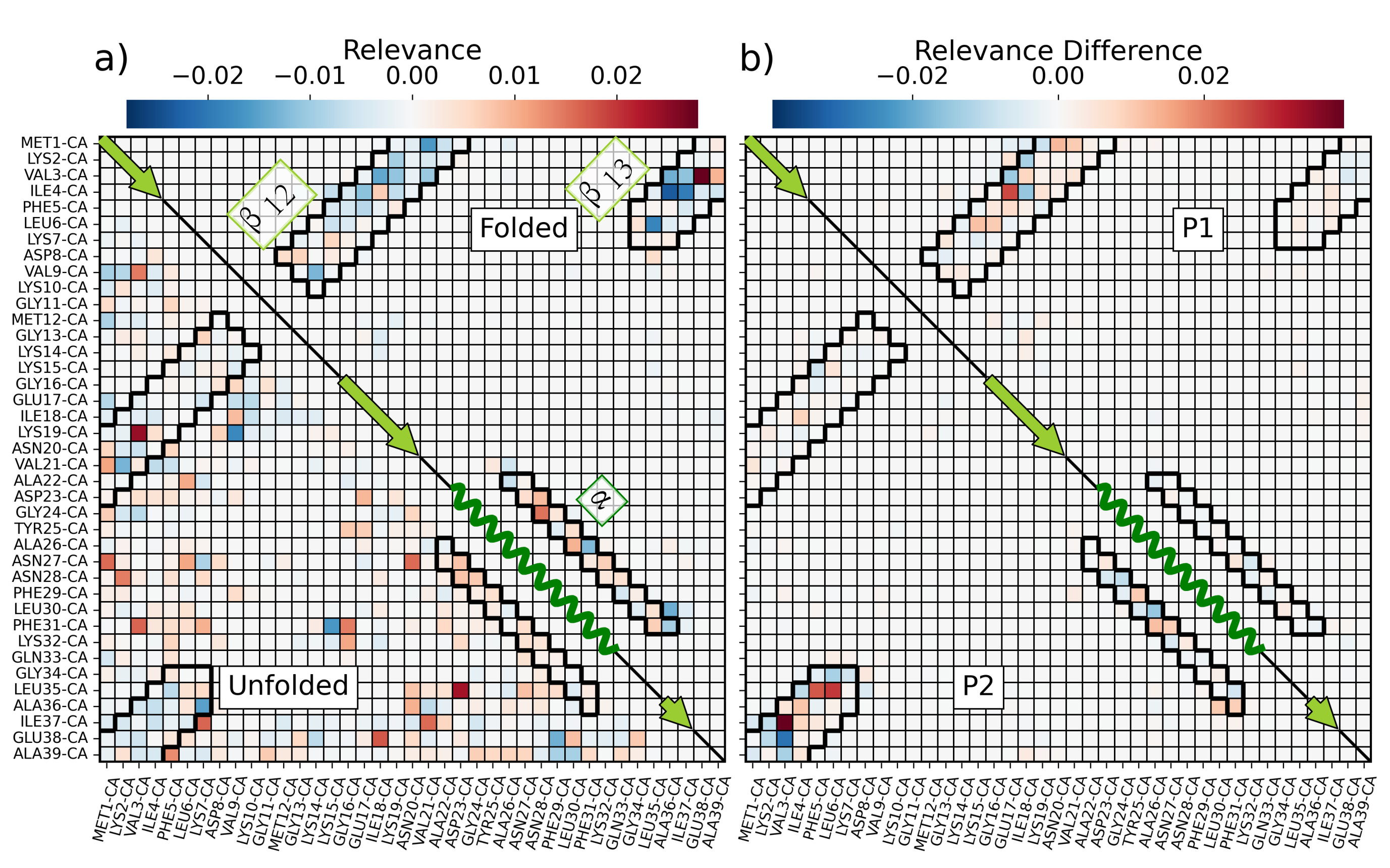}
\caption{Relevance contact maps for wild-type NTL9. a) mean relevance of amino-acid pairs in the folded state (upper right) and the unfolded state (lower left); b) mean relevance difference between the folded state and the P1 (upper right) and P2 (lower left) states, respectively, i.e. $R_{P1/P2}-R_{F}$. The regions bordered in black correspond to the contacts associated with the main secondary structure elements.}
\label{fig:ntl9_WT_contact_maps}
\end{figure}

\begin{figure}[h]
\centering
\includegraphics[width=0.8\linewidth]{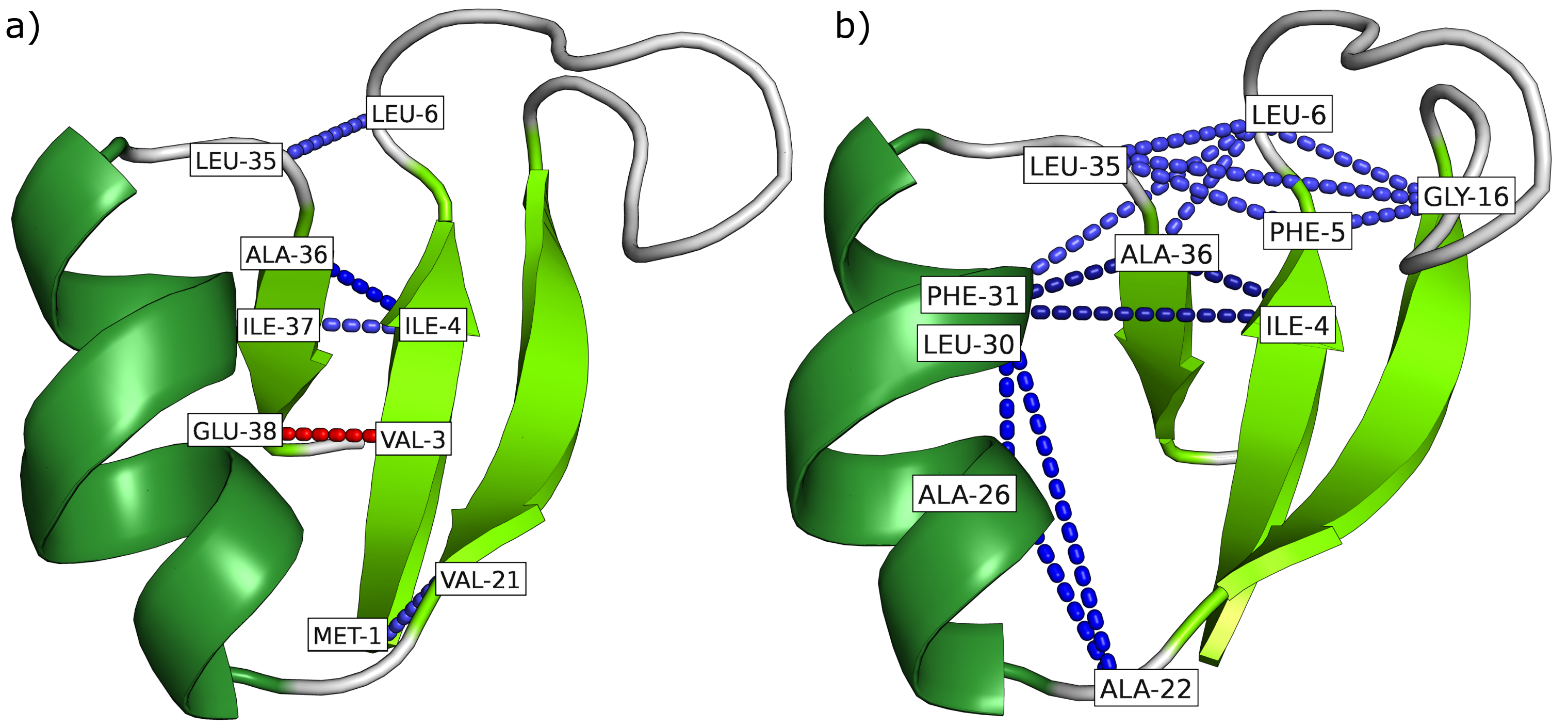}
\caption{Snapshots of folded NTL9 with the most relevant 2- and 3-body interactions learned  by the CG model. Panels a) and b) show the five most important 2- and 3-body interactions respectively. Blue lines indicate stabilizing interactions (negative relevance) and red lines destabilizing interactions (positive relevance). Darker color shades indicate stronger interaction strength (darker blue/red lines corresponds to stronger repulsive/attractive interactions). Visualizations generated with PyMOL~\cite{pymol}.}
\label{fig:ntl9_structures_2B_3B}
\end{figure}

The folded state of NTL9 contains 3 $\beta$ sheets that are formed by the residues along the C- and N-terminal regions as well as a central $\alpha$ helix (shown on the bottom left in Fig.~\ref{fig:NTL9FES}). 
The stability of this short fragment of the N-terminal domain of the Ribosomal Protein L9 is likely due to the strong hydrophobic core between the $\beta$-sheets and the $\alpha$-helix~\cite{horng_characterization_2002,anil_fine_2005}.
To test whether the model is learning these interactions, we compute the relevance score for 2- and 3-body interactions in the trained model, for structures taken from different metastable states.
The contact map in Fig.~\ref{fig:ntl9_WT_contact_maps}a) shows the 2-body relevance scores for each pair of amino-acids in the folded (upper right) and unfolded (lower left) states. 
Interestingly, the 2-body interactions stabilizing the folded state correspond to contacts associated with the main secondary structure elements, while in the unfolded state both stabilizing and destabilizing interactions are found also outside of the secondary structure. 
The strongest 2- and 3-body interactions inside the folded state are shown in Fig.~\ref{fig:ntl9_structures_2B_3B}. The strongest 2-body contributions are found in the $\beta_{13}$ sheet, most of which are stabilizing interactions. Notably, the VAL3-GLU38  interaction is destabilizing, which indicates that the CG model learns the side-chain specific interaction between the charged Glutamate and the hydrophobic Valine. 
The strongest 3-body interactions in the folded state, Fig.~\ref{fig:ntl9_structures_2B_3B} panel (b), stabilize the helix and the tertiary structure of the protein.

NTL9 can fold by two different pathways, which appear as two distinct "branches" in the free energy landscapes in Fig. \ref{fig:NTL9FES}. We examine the differences in relevance patterns between the two pathways connecting the folded to unfolded states. In panel b) of Fig.~\ref{fig:ntl9_WT_contact_maps} we show the difference in relevance for both intermediate states relative to the folded state. Here, a positive difference means that the interaction has a lower relevance in the folded state than in the intermediate state and thus that this interaction is less stable in the intermediate state. For the state indicated as P1 in Fig.~\ref{fig:NTL9FES}, many destabilizing interactions are located inside the $\beta_{12}$ sheet, whereas the difference to the folded state is essentially null in the $\beta_{13}$ sheet and in the $\alpha$-helix. This indicates that P1 corresponds to an intermediate state where the $\beta_{13}$ sheet and the $\alpha$-helix are native-like but the $\beta_{12}$ is less stable than in the native state. Indeed, analysis of the structures in P1 reveal that $\beta_{12}$ is register-shifted. The P2 state shows the opposite behavior with destabilizing interactions in the $\beta_{13}$ sheet and in the $\alpha$-helix, indicating that in the P2 state, only the $\beta_{12}$ sheet is correctly formed. 
These two folding pathways with the same characteristics are also found in previous computational studies of NTL9~\cite{snow_kinetic_2006,voelz2010molecular,lindorff-larsen_how_2011}.
Note that if the relevance attribution in the folded state shows that the network captures the interaction decay with the distance between residues, the relevance attribution in a given state provides more information than contained merely in the contact maps for these states, as can be seen by comparing the relevance attribution of both intermediate states to their distance contact maps shown in Supplementary Fig. S6.

The interpretation of the learned interactions can be pushed a step further by considering the  effects of mutations on the relevance analysis. We consider mutations of residues deemed stabilizing in the folded state.
In the machine learned C$_{\alpha}$ CG model of the protein employed here, one can straightforwardly perform a mutation by changing the  aminoacid identity, that is by changing the embedding of the corresponding C$_{\alpha}$ bead.
We select two mutations to illustrate the ability of the CG  model to learn specific interactions such as hydrophobic/hydrophilic interactions between side-chains and side-chain specific packing. In particular, the mutation ILE4ASN is chosen to disrupt the hydrophobic interaction of the $\beta$-sheets, and the mutation LEU30PHE is chosen to disrupt the central $\alpha$-helix as well as the tight packing between the $\alpha$-helix and the $\beta$-sheets.
These residues are flagged as important in the analysis above and have been shown in mutation experiments to play a role in the stabilization of the folded state~\cite{anil_fine_2005,sato2017}.

\begin{figure}[h]
\centering
\includegraphics[width=1.0\linewidth]{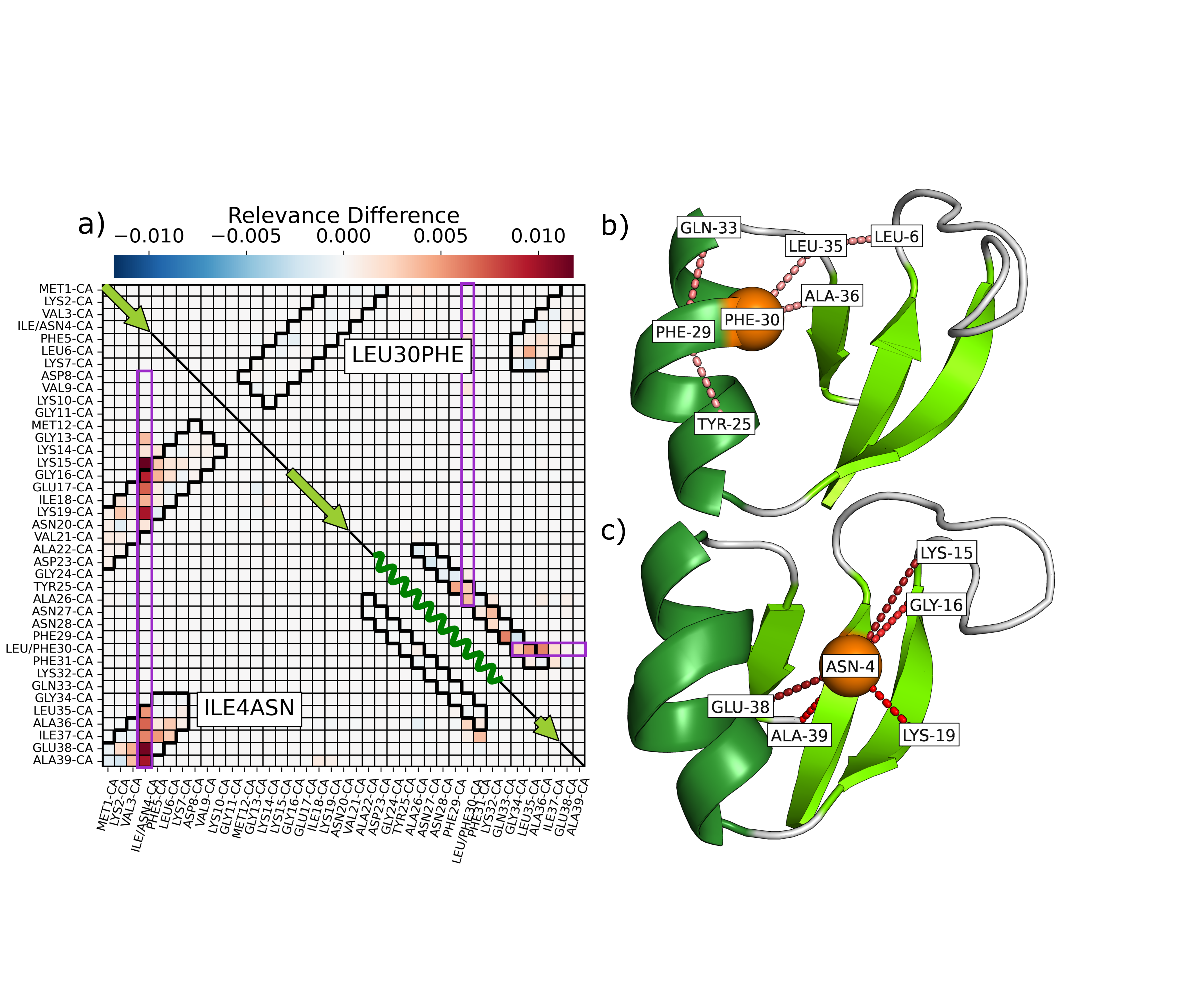}
\caption{Effect of mutations on the model prediction. Panel a) shows the mean relevance difference for each amino-acid pair to the wild-type prediction, i.e. $R_{mut}-R_{WT}$. The upper right half corresponds to the LEU30PHE and the lower left to the ILE4ASN mutation. A red (positive) interaction corresponds to a higher relevance in the mutated state than in the wild-type, thus a destabilized interaction. Panels b) and c) show an example structure with the mutated residue highlighted in orange as well as strongest interactions flagged by the network interpretation in the same fashion as in Fig.~\ref{fig:ntl9_structures_2B_3B}.}
\label{fig:ntl9_mutations}
\end{figure}

In Fig.~\ref{fig:ntl9_mutations}a) we show the 2-body relevance difference between the mutated states (LEU30PHE on the upper right and ILE4ASN on the lower left half) and the wild-type folded state. Replacing the identities of hydrophobic Isoleucine by polar Asparagine of about the same size at position 4 introduces a strong destabilization of all contacts with hydrophobic residues in the neighboring sheets, as it is also visualized in panel c). In the crystal structure of NTL9, LEU30 is tightly packed between the $\alpha$-helix and the $\beta$-sheets and mutation studies suggest that even a small change in side-chain size has a destabilizing effect~\cite{anil_fine_2005}. Indeed, replacing the identity of Leucine 30 by the bigger Phenylalanine in our CG model induces a destabilization of the entire $\alpha$-helix (see Fig.~\ref{fig:ntl9_mutations}, panels a) and b)). Interestingly, the disruption also has an effect on contacts inside the $\beta_{13}$ sheet that do not directly involve the mutated residue, indicating that the model has indeed learned non-trivial many-body interactions.
These findings further corroborate the ability of the CG model to learn amino-acid specific interactions in a $C_{\alpha}$-only representation.

\section*{Discussion}

In this work, we propose an extension of GNN-LRP to interpret the effective energy of machine-learned CG protein models in terms of a multi-body decomposition. We have shown on the application to CG fluids that the learned interactions are physically meaningful and consistent even if different ML architectures are used. 
The explanations provided by GNN-LRP indicate when multi-body interactions are required to recover the thermodynamics of the fine-grained system, showing the suitability of this higher-order explanation method to ML CG force-fields. Moreover, the multi-body relevance contributions show that the different ML models have learned similar physically relevant interatomic interactions, indicating that these models effectively learn the same underlying potential energy surface. 
The application of this idea on ML CG protein models allows to disentangle the strength of the different interactions between residues. It can also be used to evaluate the effect of mutations in the model and shows that bottom-up $C_{\alpha}$-based ML CG models capture amino-acid specific interactions without explicit representation of the side-chains.

These results provide reassurance that bottom-up machine learned CG force-fields indeed learn physically relevant terms by approximating the many-body potential of mean-force associated with the integration of part the degrees of freedom~\cite{noid2008mscg1}.
We note that the methods introduced are model agnostic and can be used in general to interpret machine learned potentials of different systems at different resolutions.

Future work is still needed to refine these concepts to provide greater insight into ML models and allow researchers to be more systematic in their choice of architecture and functionalization. 
It is our hope that this work helps lay the groundwork for researchers to better understand the outputs of their models as well as give the coarse-graining community a way to probe these learned many-body effects more explicitly. 

\section*{Methods}

\subsection*{Machine learned CG Potentials}
Graph Neural Networks (GNNs) have been proposed as a promising method to learn interatomic potentials~\cite{gilmer2017neural}, and many different model architectures have been developed in recent years~\cite{unke2021review}.
GNNs represent the underlying atomic details using structured graph data where nodes represent atoms and use the idea that locality dominates the energy landscape to draw edges between nodes if two nodes are within a pre-defined cutoff distance.
In GNNs there are generally three steps that go into producing the network output based on the input positions and node identities: (i) a message-passing step, where neighboring nodes (connected by edges) exchange information about their respective feature values, (ii) an update step, where the node features are modified based on the received messages, and (iii) a final readout step, where the features of each node are used to predict the target property~\cite{gilmer2017neural}.
Once the node features are fed through the readout layer the network can make use of backpropagation to extract the force by taking derivatives of the energy based on molecule positions that can then be used to train during force matching or to propagate the dynamics.

In this manuscript, we examine two different GNN architectures for the  effective CG energy: PaiNN~\cite{schutt2021equivariant} and SO3Net~\cite{schutt2023schnetpack}, two equivariant message-passing architectures that mainly differ on the order of their SO3-equivariant features ($l_\mathrm{max}=1$ for PaiNN and $l_\mathrm{max}=2$ for SO3Net).
Both preserve the basic euclidean symmetries of the system, notably those of translation, rotation and reflection. Both architectures are parametrized to reproduce the CG potential of mean force using the force-matching approach~\cite{ercolessi_interatomic_1994, izvekov_multiscale_2005, noid2008mscg1, wang_machine_2019,husic2020coarse}.
As a comparison to more classical methods, Inverse Monte Carlo (IMC)~\cite{lyubartsev_calculation_1995,lyubartsev_systematic_2009} is also performed using the votca library~\cite{ruhle2009versatile} with results shown in the Supplementary Section S3.

\subsection*{Layerwise-relevance propagation}
Layerwise-relevance propagation (LRP) has emerged as a method to explain model predictions in a post-hoc and model agnostic manner~\cite{montavon2019layer, bach2015pixel, montavon2017explaining, montavon2018methods}. Originally, LRP has been used to obtain first-order explanations in the form of relevance attributions (also referred to as relevance scores) in the input domain. E.g., for image classification tasks, the relevance attributions would indicate to what extent a respective pixel is responsible for the network decision~\cite{bach2015pixel, lapuschkin2019unmasking}. The relevance attributions can be visualized in the input domain in form of a heat map, where large relevance attributions highlight features of the classified object that predominantly lead to the respective decision of the neural network~\cite{bach2015pixel}. Recently, efforts have been made to adapt LRP and other explanation methods to regression tasks~\cite{letzgus2022toward, letzgus2024xpertai, spooner2021counterfactual}, 
such as, e.g., the prediction of atomization energies~\cite{letzgus2022toward, schnake2022gnnlrp}.

Pixel-wise relevance attributions have contributed enormously to a better understanding of the inner workings of ANNs. 
However, in some cases, restricting explanations to first-order (input features) may result in oversimplified explanations. Especially for problems where the interaction between several input nodes is considerably strong, the relevance information of higher-order features becomes increasingly important. This is the case for the coarse-grained systems considered in this study, where multi-body interactions are essential~\cite{wang_comparative_2009,larini_coarse-grained_2012,wang2021multi, zaporozhets2023multibody}. To this end, a variety of higher-order explanation frameworks have been introduced~\cite{eberle2020building, ying2019gnnexplainer, schnake2022gnnlrp,xiong2022efficient, blucher2022preddiff, faber2021comparing, janizek2021explaining, lundberg2020local, dhamdhere2019shapley}. One of those methods is GNN-LRP, which extends LRP to higher-order explanations for GNNs~\cite{schnake2022gnnlrp, xiong2022efficient}. In the following, we will give a brief summary of the methodology of GNN-LRP, first describing first-order relevance propagation and then extending it to higher-order. For an in-depth introduction, please refer to~\cite{montavon2019layer, schnake2022gnnlrp}.

In general, relevance attributions are obtained by propagating relevance from the model output back to the input features, with the relevance being a conserved quantity much the same as the mass flow through a pipe or electrical current through a junction. This means that the sum relevance attribution of a layer in the network remains the same as it is propagated backwards through the network. The relevance propagation from the neurons $\{j\}$ to a lower-layer neuron $i$ reads
\begin{equation}
R_i^l = \sum_j \frac{q_{ij}}{\sum_i q_{ij}} \cdot R_j^{l+1}~,
\label{eq:first_order_lrp}
\end{equation}
where $q_{ij}$ quantifies the contribution of neuron $i$ to the activation of neuron $j$. 
This propagation approach can be applied subsequently from the output neuron until all relevance is attributed to the input neurons. There are multiple ways to define $q_{ij}$ associated with different propagation rules (see also \cite{montavon2019layer}). According to the deep Taylor decomposition~\cite{montavon2017explaining}, LRP is equivalent to a decomposition of the neural network into several first-order Taylor expansions. In this picture, the different LRP rules implicitly help find suitable root points for the respective Taylor expansions~\cite{montavon2017explaining}.

In contrast to the first-order node-wise relevance attributions, GNN-LRP explains the network prediction based on so-called relevance ``walks". Each walk is a collection of connected edges and nodes of the input graph. The length of the walks is dictated by the number of interaction layers (message passes) in the model. Hence, e.g., a model with two interaction layers would allow for up to 3-body walks. The concept of GNN-LRP is illustrated in Fig.~\ref{fig:LRPUnderstanding}a) for a model with two interaction layers applied to an exemplary system composed of four particles (e.g., CG beads). The top part of the figure shows the forward pass of a common GNN starting from the feature embedding on the input graph until the model output, while the bottom figure illustrates how the walks with their associated relevance attributions are constructed. Note that the connections for message passing in the GNN are defined by the graph structure. Depending on the number of interaction layers and the size of the node features, each graph node will have multiple corresponding neurons in the GNN. We distinguish between graph nodes ($i, j, k, ...$) and neurons ($a, b, c, ...$) by using subscript and superscript, respectively. Similar to Eq.~\ref{eq:first_order_lrp}, the relevance of a walk  $\mathcal{W}=[j, k, l]$ is obtained by propagating back the relevance from upper-layer neurons
\begin{equation}
    R_{jkl}^b = \sum_{c} \frac{\lambda_{jk}Q_{jk}^{bc} }{\sum_{b,j}\lambda_{jk}Q_{jk}^{bc}}  R_{kl}^c
\label{eq:gnn_lrp_bp}
\end{equation}
where $\lambda_{ij}$ denotes the edge feature between graph nodes $i$ and $j$, and $Q_{jk}^{bc}$ is the contribution of node $j$ with its respective neuron $b$ to neuron $c$ with the associated node $k$. Depending on the propagation rule, again, $Q_{jk}^{bc}$ takes different forms. For more information on how $Q_{jk}^{bc}$ is obtained considering the propagation rules used in this work, please refer to the Supplementary Section S5. Note that Eq.~\ref{eq:gnn_lrp_bp} yields the walk relevance for a specific neuron $b$. The total relevance of the respective walk is given by summing over all corresponding neurons:
\begin{equation}
    R_{jkl} = \sum_b R_{jkl}^b~.
\end{equation}



The relevance of each walk can be seen as a contribution to the model output resulting from interactions between the subset of particles in the graph corresponding to the respective walk. More precisely, the relevance of a walk is associated with a sequence of interactions between pairs of particles or a single bead with itself. However, such a sequence of interactions is rather abstract and cannot be associated with a physical quantity in a meaningful way. In order to obtain a quantity that we can interpret as a multi-body decomposition of the output, we aggregate a collection of walks to so-called $n$-body contributions, as depicted in Fig.~\ref{fig:LRPUnderstanding}b). As suggested by Schnake~et.~al.~\cite{schnake2022gnnlrp}, this can be achieved by summing up the relevance of all walks $\left\{\mathcal{W}\right\}$ that are part of a certain graph substructure $\mathcal{S}$. Hence, the relevance of a substructure $\mathcal{S}$ is given as
\begin{align}\label{eq:graph_rel_def}
    R_{\mathcal{S}} = \sum_{\mathcal{W} \in \mathcal{S}} R_\mathcal{W}~.
\end{align}
In this way, for the example of Fig.~\ref{fig:LRPUnderstanding}, we obtain 2-body and 3-body relevance attributions by summing over the relevance attributions associated to the walks which are part of the respective 2-body or 3-body substructure. In the case of energy predictions, this yields the energy contribution of the respective $n$-body substructures.


\section*{Acknowledgements}
{We gratefully acknowledge funding from the Deutsche Forschungsgemeinschaft
DFG (SFB/TRR 186, Project A12; SFB 1114, Projects B03, B08, and A04; SFB 1078, Project C7), the National Science Foundation (PHY-2019745), and the Einstein Foundation Berlin (Project 0420815101), and the computing time provided on the supercomputer Lise at NHR@ZIB as part of the NHR infrastructure. K.R.M. was in part supported by the German Ministry for Education and Research (BMBF) under Grants 01IS14013A-E, 01GQ1115, 01GQ0850, 01IS18025A, 031L0207D, and 01IS18037A, and by the Institute of Information \& Communications Technology Planning \& Evaluation (IITP) grants funded by the Korean government (MSIT) No. 2019-0-00079, Artificial Intelligence Graduate School Program, Korea University and No. 2022-0-00984, Development of Artificial Intelligence Technology for Personalized Plug-and-Play Explanation and Verification of Explanation. 
}

\section*{Data Availability}
Simulation data and the code to reproduce the analysis and the plots shown in the manuscript are accessible at 
\url{https://box.fu-berlin.de/apps/files/?dir=/peering_inside_the_black_box_supplementary_data_and_code&fileid=555049372}


\newpage

\bibliographystyle{naturemag}

\begin{thebibliography}{10}
\expandafter\ifx\csname url\endcsname\relax
  \def\url#1{\texttt{#1}}\fi
\expandafter\ifx\csname urlprefix\endcsname\relax\def\urlprefix{URL }\fi
\providecommand{\bibinfo}[2]{#2}
\providecommand{\eprint}[2][]{\url{#2}}

\bibitem{best_atomistic_2019}
\bibinfo{author}{Best, R.~B.}
\newblock \bibinfo{title}{Atomistic force fields for proteins}.
\newblock In \bibinfo{editor}{Bonomi, M.} \& \bibinfo{editor}{Camilloni, C.}
  (eds.) \emph{\bibinfo{booktitle}{Biomolecular Simulations: Methods and
  Protocols}}, \bibinfo{pages}{3--19} (\bibinfo{publisher}{Springer},
  \bibinfo{year}{2019}).

\bibitem{monticelli2013force}
\bibinfo{author}{Monticelli, L.} \& \bibinfo{author}{Tieleman, D.~P.}
\newblock \bibinfo{title}{Force fields for classical molecular dynamics}.
\newblock In \bibinfo{editor}{Monticelli, L.} \& \bibinfo{editor}{Salonen, E.}
  (eds.) \emph{\bibinfo{booktitle}{Biomolecular Simulations: Methods and
  Protocols}}, \bibinfo{pages}{197--213} (\bibinfo{publisher}{Humana Press},
  \bibinfo{year}{2013}).

\bibitem{lindorff-larsen_systematic_2012}
\bibinfo{author}{Lindorff-Larsen, K.} \emph{et~al.}
\newblock \bibinfo{title}{Systematic validation of protein force fields against
  experimental data}.
\newblock \emph{\bibinfo{journal}{{PloS} one}} \textbf{\bibinfo{volume}{7}},
  \bibinfo{pages}{e32131} (\bibinfo{year}{2012}).

\bibitem{robustelli_developing_2018}
\bibinfo{author}{Robustelli, P.}, \bibinfo{author}{Piana, S.} \&
  \bibinfo{author}{Shaw, D.~E.}
\newblock \bibinfo{title}{Developing a molecular dynamics force field for both
  folded and disordered protein states}.
\newblock \emph{\bibinfo{journal}{Proc. Natl. Acad. Sci. USA}}
  \textbf{\bibinfo{volume}{115}}, \bibinfo{pages}{E4758--E4766}
  (\bibinfo{year}{2018}).

\bibitem{behler_generalized_2007}
\bibinfo{author}{Behler, J.} \& \bibinfo{author}{Parrinello, M.}
\newblock \bibinfo{title}{Generalized neural-network representation of
  high-dimensional potential-energy surfaces}.
\newblock \emph{\bibinfo{journal}{Phys. Rev. Lett.}}
  \textbf{\bibinfo{volume}{98}}, \bibinfo{pages}{146401}
  (\bibinfo{year}{2007}).

\bibitem{noe_machine_2020}
\bibinfo{author}{No{\'e}, F.}, \bibinfo{author}{Tkatchenko, A.},
  \bibinfo{author}{M{\"u}ller, K.-R.} \& \bibinfo{author}{Clementi, C.}
\newblock \bibinfo{title}{Machine learning for molecular simulation}.
\newblock \emph{\bibinfo{journal}{Annu. Rev. Phys. Chem.}}
  \textbf{\bibinfo{volume}{71}}, \bibinfo{pages}{361--390}
  (\bibinfo{year}{2020}).

\bibitem{scarselli_graph_2009}
\bibinfo{author}{Scarselli, F.}, \bibinfo{author}{Gori, M.},
  \bibinfo{author}{Tsoi, A.~C.}, \bibinfo{author}{Hagenbuchner, M.} \&
  \bibinfo{author}{Monfardini, G.}
\newblock \bibinfo{title}{The graph neural network model}.
\newblock \emph{\bibinfo{journal}{IEEE Trans. Neural Netw. Learn. Syst.}}
  \textbf{\bibinfo{volume}{20}}, \bibinfo{pages}{61--80}
  (\bibinfo{year}{2009}).

\bibitem{duvenaud_convolutional_2015}
\bibinfo{author}{Duvenaud, D.~K.} \emph{et~al.}
\newblock \bibinfo{title}{Convolutional networks on graphs for learning
  molecular fingerprints}.
\newblock In \emph{\bibinfo{booktitle}{Adv. Neural Inf. Process. Syst.}},
  vol.~\bibinfo{volume}{28} (\bibinfo{publisher}{Curran Associates, Inc.},
  \bibinfo{year}{2015}).

\bibitem{kearnes_molecular_2016}
\bibinfo{author}{Kearnes, S.}, \bibinfo{author}{{McCloskey}, K.},
  \bibinfo{author}{Berndl, M.}, \bibinfo{author}{Pande, V.} \&
  \bibinfo{author}{Riley, P.}
\newblock \bibinfo{title}{Molecular graph convolutions: moving beyond
  fingerprints}.
\newblock \emph{\bibinfo{journal}{J. Comput. Aided Mol. Des.}}
  \textbf{\bibinfo{volume}{30}}, \bibinfo{pages}{595--608}
  (\bibinfo{year}{2016}).

\bibitem{gilmer2017neural}
\bibinfo{author}{Gilmer, J.}, \bibinfo{author}{Schoenholz, S.~S.},
  \bibinfo{author}{Riley, P.~F.}, \bibinfo{author}{Vinyals, O.} \&
  \bibinfo{author}{Dahl, G.~E.}
\newblock \bibinfo{title}{Neural message passing for quantum chemistry}.
\newblock In \emph{\bibinfo{booktitle}{Int. Conf. Mach. Learn.}},
  \bibinfo{pages}{1263--1272} (\bibinfo{organization}{PMLR},
  \bibinfo{year}{2017}).

\bibitem{schuett2017dtnn}
\bibinfo{author}{Sch{\"u}tt, K.~T.}, \bibinfo{author}{Arbabzadah, F.},
  \bibinfo{author}{Chmiela, S.}, \bibinfo{author}{M{\"u}ller, K.~R.} \&
  \bibinfo{author}{Tkatchenko, A.}
\newblock \bibinfo{title}{Quantum-chemical insights from deep tensor neural
  networks}.
\newblock \emph{\bibinfo{journal}{Nat. Commun.}} \textbf{\bibinfo{volume}{8}},
  \bibinfo{pages}{13890} (\bibinfo{year}{2017}).

\bibitem{schutt_schnet_2018}
\bibinfo{author}{Sch{\"u}tt, K.~T.}, \bibinfo{author}{Sauceda, H.~E.},
  \bibinfo{author}{Kindermans, P.-J.}, \bibinfo{author}{Tkatchenko, A.} \&
  \bibinfo{author}{M{\"u}ller, K.-R.}
\newblock \bibinfo{title}{{SchNet} - a deep learning architecture for molecules
  and materials}.
\newblock \emph{\bibinfo{journal}{J. Chem. Phys.}}
  \textbf{\bibinfo{volume}{148}}, \bibinfo{pages}{241722}
  (\bibinfo{year}{2018}).

\bibitem{unke2019physnet}
\bibinfo{author}{Unke, O.~T.} \& \bibinfo{author}{Meuwly, M.}
\newblock \bibinfo{title}{{PhysNet}: A neural network for predicting energies,
  forces, dipole moments, and partial charges}.
\newblock \emph{\bibinfo{journal}{J. Chem. Theory Comput.}}
  \textbf{\bibinfo{volume}{15}}, \bibinfo{pages}{3678--3693}
  (\bibinfo{year}{2019}).

\bibitem{Unke2024}
\bibinfo{author}{Unke, O.~T.} \emph{et~al.}
\newblock \bibinfo{title}{Biomolecular dynamics with machine-learned
  quantum-mechanical force fields trained on diverse chemical fragments}.
\newblock \emph{\bibinfo{journal}{Sci. Adv.}} \textbf{\bibinfo{volume}{10}},
  \bibinfo{pages}{eadn4397} (\bibinfo{year}{2024}).

\bibitem{ceriotti2022beyond}
\bibinfo{author}{Ceriotti, M.}
\newblock \bibinfo{title}{Beyond potentials: Integrated machine learning models
  for materials}.
\newblock \emph{\bibinfo{journal}{Mrs Bulletin}} \textbf{\bibinfo{volume}{47}},
  \bibinfo{pages}{1045--1053} (\bibinfo{year}{2022}).

\bibitem{samek21xaireview}
\bibinfo{author}{Samek, W.}, \bibinfo{author}{Montavon, G.},
  \bibinfo{author}{Lapuschkin, S.}, \bibinfo{author}{Anders, C.~J.} \&
  \bibinfo{author}{M{\"u}ller, K.-R.}
\newblock \bibinfo{title}{Explaining deep neural networks and beyond: A review
  of methods and applications}.
\newblock \emph{\bibinfo{journal}{Proc. IEEE}} \textbf{\bibinfo{volume}{109}},
  \bibinfo{pages}{247--278} (\bibinfo{year}{2021}).

\bibitem{schuett2019xaiqc}
\bibinfo{author}{Sch{\"u}tt, K.~T.}, \bibinfo{author}{Gastegger, M.},
  \bibinfo{author}{Tkatchenko, A.} \& \bibinfo{author}{M{\"u}ller, K.-R.}
\newblock \emph{\bibinfo{title}{Quantum-Chemical Insights from Interpretable
  Atomistic Neural Networks}}, \bibinfo{pages}{311--330}
  (\bibinfo{publisher}{Springer International Publishing},
  \bibinfo{address}{Cham}, \bibinfo{year}{2019}).

\bibitem{gallegos_explainable_2024}
\bibinfo{author}{Gallegos, M.}, \bibinfo{author}{Vassilev-Galindo, V.},
  \bibinfo{author}{Poltavsky, I.}, \bibinfo{author}{Mart{\'i}n~Pend{\'a}s,
  {\'A}.} \& \bibinfo{author}{Tkatchenko, A.}
\newblock \bibinfo{title}{Explainable chemical artificial intelligence from
  accurate machine learning of real-space chemical descriptors}.
\newblock \emph{\bibinfo{journal}{Nat. Commun.}} \textbf{\bibinfo{volume}{15}},
  \bibinfo{pages}{4345} (\bibinfo{year}{2024}).

\bibitem{zhang2022protgnn}
\bibinfo{author}{Zhang, Z.}, \bibinfo{author}{Liu, Q.}, \bibinfo{author}{Wang,
  H.}, \bibinfo{author}{Lu, C.} \& \bibinfo{author}{Lee, C.}
\newblock \bibinfo{title}{{ProtGNN}: Towards self-explaining graph neural
  networks}.
\newblock In \emph{\bibinfo{booktitle}{Proc. AAAI Conf. Artif. Intell.}},
  vol.~\bibinfo{volume}{36}, \bibinfo{pages}{9127--9135}
  (\bibinfo{year}{2022}).

\bibitem{lundberg17unified}
\bibinfo{author}{Lundberg, S.~M.} \& \bibinfo{author}{Lee, S.-I.}
\newblock \bibinfo{title}{A unified approach to interpreting model
  predictions}.
\newblock In \bibinfo{editor}{Guyon, I.} \emph{et~al.} (eds.)
  \emph{\bibinfo{booktitle}{Adv. Neural Inf. Process. Syst.}},
  vol.~\bibinfo{volume}{30} (\bibinfo{publisher}{Curran Associates, Inc.},
  \bibinfo{year}{2017}).

\bibitem{bach2015pixel}
\bibinfo{author}{Bach, S.} \emph{et~al.}
\newblock \bibinfo{title}{On pixel-wise explanations for non-linear classifier
  decisions by layer-wise relevance propagation}.
\newblock \emph{\bibinfo{journal}{PloS one}} \textbf{\bibinfo{volume}{10}},
  \bibinfo{pages}{e0130140} (\bibinfo{year}{2015}).

\bibitem{sundararajan2017axiomatic}
\bibinfo{author}{Sundararajan, M.}, \bibinfo{author}{Taly, A.} \&
  \bibinfo{author}{Yan, Q.}
\newblock \bibinfo{title}{Axiomatic attribution for deep networks}.
\newblock In \emph{\bibinfo{booktitle}{Int. Conf. Mach. Learn.}},
  \bibinfo{pages}{3319--3328} (\bibinfo{organization}{PMLR},
  \bibinfo{year}{2017}).

\bibitem{ribeiro2016should}
\bibinfo{author}{Ribeiro, M.~T.}, \bibinfo{author}{Singh, S.} \&
  \bibinfo{author}{Guestrin, C.}
\newblock \bibinfo{title}{"{Why should I trust you?}" explaining the
  predictions of any classifier}.
\newblock In \emph{\bibinfo{booktitle}{Proceedings of the 22nd ACM SIGKDD
  international conference on knowledge discovery and data mining}},
  \bibinfo{pages}{1135--1144} (\bibinfo{year}{2016}).

\bibitem{debnath1991structure}
\bibinfo{author}{Debnath, A.~K.}, \bibinfo{author}{Lopez~de Compadre, R.~L.},
  \bibinfo{author}{Debnath, G.}, \bibinfo{author}{Shusterman, A.~J.} \&
  \bibinfo{author}{Hansch, C.}
\newblock \bibinfo{title}{Structure-activity relationship of mutagenic aromatic
  and heteroaromatic nitro compounds. correlation with molecular orbital
  energies and hydrophobicity}.
\newblock \emph{\bibinfo{journal}{J. Med. Chem.}}
  \textbf{\bibinfo{volume}{34}}, \bibinfo{pages}{786--797}
  (\bibinfo{year}{1991}).

\bibitem{Kazius2005DerivMutag}
\bibinfo{author}{Kazius, J.}, \bibinfo{author}{McGuire, R.} \&
  \bibinfo{author}{Bursi, R.}
\newblock \bibinfo{title}{Derivation and validation of toxicophores for
  mutagenicity prediction}.
\newblock \emph{\bibinfo{journal}{J. Med. Chem.}}
  \textbf{\bibinfo{volume}{48}}, \bibinfo{pages}{312--320}
  (\bibinfo{year}{2005}).

\bibitem{baehrens2010explain}
\bibinfo{author}{Baehrens, D.} \emph{et~al.}
\newblock \bibinfo{title}{How to explain individual classification decisions}.
\newblock \emph{\bibinfo{journal}{J. Mach. Learn. Res.}}
  \textbf{\bibinfo{volume}{11}}, \bibinfo{pages}{1803--1831}
  (\bibinfo{year}{2010}).

\bibitem{letzgus2022toward}
\bibinfo{author}{Letzgus, S.} \emph{et~al.}
\newblock \bibinfo{title}{Toward explainable artificial intelligence for
  regression models: A methodological perspective}.
\newblock \emph{\bibinfo{journal}{IEEE Signal Processing Magazine}}
  \textbf{\bibinfo{volume}{39}}, \bibinfo{pages}{40--58}
  (\bibinfo{year}{2022}).

\bibitem{schnake2022gnnlrp}
\bibinfo{author}{Schnake, T.} \emph{et~al.}
\newblock \bibinfo{title}{Higher-order explanations of graph neural networks
  via relevant walks}.
\newblock \emph{\bibinfo{journal}{IEEE Trans Pattern Anal Mach Intell}}
  \textbf{\bibinfo{volume}{44}}, \bibinfo{pages}{7581--7596}
  (\bibinfo{year}{2022}).

\bibitem{jose20drugdisc}
\bibinfo{author}{Jim{\'e}nez-Luna, J.}, \bibinfo{author}{Grisoni, F.} \&
  \bibinfo{author}{Schneider, G.}
\newblock \bibinfo{title}{Drug discovery with explainable artificial
  intelligence}.
\newblock \emph{\bibinfo{journal}{Nat. Mach. Intell.}}
  \textbf{\bibinfo{volume}{2}}, \bibinfo{pages}{573--584}
  (\bibinfo{year}{2020}).

\bibitem{wellawatte2023perspective}
\bibinfo{author}{Wellawatte, G.~P.}, \bibinfo{author}{Gandhi, H.~A.},
  \bibinfo{author}{Seshadri, A.} \& \bibinfo{author}{White, A.~D.}
\newblock \bibinfo{title}{A perspective on explanations of molecular prediction
  models}.
\newblock \emph{\bibinfo{journal}{J. Chem. Theory Comput.}}
  \textbf{\bibinfo{volume}{19}}, \bibinfo{pages}{2149--2160}
  (\bibinfo{year}{2023}).

\bibitem{mcclo19usingatt}
\bibinfo{author}{McCloskey, K.}, \bibinfo{author}{Taly, A.},
  \bibinfo{author}{Monti, F.}, \bibinfo{author}{Brenner, M.~P.} \&
  \bibinfo{author}{Colwell, L.~J.}
\newblock \bibinfo{title}{Using attribution to decode binding mechanism in
  neural network models for chemistry}.
\newblock \emph{\bibinfo{journal}{Proc. Natl. Acad. Sci. USA}}
  \textbf{\bibinfo{volume}{116}}, \bibinfo{pages}{11624--11629}
  (\bibinfo{year}{2019}).

\bibitem{lill2019elucidating}
\bibinfo{author}{Mahmoud, A.~H.}, \bibinfo{author}{Masters, M.~R.},
  \bibinfo{author}{Yang, Y.} \& \bibinfo{author}{Lill, M.~A.}
\newblock \bibinfo{title}{Elucidating the multiple roles of hydration for
  accurate protein-ligand binding prediction via deep learning}.
\newblock \emph{\bibinfo{journal}{Commun. Chem.}} \textbf{\bibinfo{volume}{3}},
  \bibinfo{pages}{1--13} (\bibinfo{year}{2020}).

\bibitem{hochuli2018visualizing}
\bibinfo{author}{Hochuli, J.}, \bibinfo{author}{Helbling, A.},
  \bibinfo{author}{Skaist, T.}, \bibinfo{author}{Ragoza, M.} \&
  \bibinfo{author}{Koes, D.~R.}
\newblock \bibinfo{title}{Visualizing convolutional neural network
  protein-ligand scoring}.
\newblock \emph{\bibinfo{journal}{J. Mol. Graph. Model.}}
  \textbf{\bibinfo{volume}{84}}, \bibinfo{pages}{96--108}
  (\bibinfo{year}{2018}).

\bibitem{yang2023explainable}
\bibinfo{author}{Yang, C.-I.} \& \bibinfo{author}{Li, Y.-P.}
\newblock \bibinfo{title}{Explainable uncertainty quantifications for deep
  learning-based molecular property prediction}.
\newblock \emph{\bibinfo{journal}{J. Cheminform.}}
  \textbf{\bibinfo{volume}{15}}, \bibinfo{pages}{13} (\bibinfo{year}{2023}).

\bibitem{roy2023learning}
\bibinfo{author}{Roy, S.}, \bibinfo{author}{D{\"u}rholt, J.~P.},
  \bibinfo{author}{Asche, T.~S.}, \bibinfo{author}{Zipoli, F.} \&
  \bibinfo{author}{G{\'o}mez-Bombarelli, R.}
\newblock \bibinfo{title}{Learning a reactive potential for silica-water
  through uncertainty attribution} (\bibinfo{year}{2023}).
\newblock \bibinfo{note}{Preprint at \url{https://arxiv.org/abs/2307.01705}}.

\bibitem{durumeric2023using}
\bibinfo{author}{Durumeric, A.~E.} \& \bibinfo{author}{Voth, G.~A.}
\newblock \bibinfo{title}{Using classifiers to understand coarse-grained models
  and their fidelity with the underlying all-atom systems}.
\newblock \emph{\bibinfo{journal}{J. Chem. Phys.}}
  \textbf{\bibinfo{volume}{158}} (\bibinfo{year}{2023}).

\bibitem{jiang2017coarse}
\bibinfo{author}{Jiang, C.}, \bibinfo{author}{Ouyang, J.},
  \bibinfo{author}{Wang, L.}, \bibinfo{author}{Liu, Q.} \& \bibinfo{author}{Li,
  W.}
\newblock \bibinfo{title}{Coarse graining of the fully atomic methane models to
  monatomic isotropic models using relative entropy minimization}.
\newblock \emph{\bibinfo{journal}{J. Mol. Liq.}}
  \textbf{\bibinfo{volume}{242}}, \bibinfo{pages}{1138--1147}
  (\bibinfo{year}{2017}).

\bibitem{zhang_deepcg_2018}
\bibinfo{author}{Zhang, L.}, \bibinfo{author}{Han, J.}, \bibinfo{author}{Wang,
  H.}, \bibinfo{author}{Car, R.} \& \bibinfo{author}{E, W.}
\newblock \bibinfo{title}{{DeePCG}: Constructing coarse-grained models via deep
  neural networks}.
\newblock \emph{\bibinfo{journal}{J. Chem. Phys.}}
  \textbf{\bibinfo{volume}{149}}, \bibinfo{pages}{034101}
  (\bibinfo{year}{2018}).

\bibitem{husic2020coarse}
\bibinfo{author}{Husic, B.~E.} \emph{et~al.}
\newblock \bibinfo{title}{Coarse graining molecular dynamics with graph neural
  networks}.
\newblock \emph{\bibinfo{journal}{J. Chem. Phys.}}
  \textbf{\bibinfo{volume}{153}} (\bibinfo{year}{2020}).

\bibitem{majewski2023machine}
\bibinfo{author}{Majewski, M.} \emph{et~al.}
\newblock \bibinfo{title}{Machine learning coarse-grained potentials of protein
  thermodynamics}.
\newblock \emph{\bibinfo{journal}{Nat. Commun.}} \textbf{\bibinfo{volume}{14}},
  \bibinfo{pages}{5739} (\bibinfo{year}{2023}).

\bibitem{charron2023}
\bibinfo{author}{Charron, N.~E.} \emph{et~al.}
\newblock \bibinfo{title}{Navigating protein landscapes with a machine-learned
  transferable coarse-grained model} (\bibinfo{year}{2023}).
\newblock \bibinfo{note}{Preprint at \url{https://arxiv.org/abs/2310.18278}}.

\bibitem{zaporozhets2023multibody}
\bibinfo{author}{Zaporozhets, I.} \& \bibinfo{author}{Clementi, C.}
\newblock \bibinfo{title}{Multibody terms in protein coarse-grained models: A
  top-down perspective}.
\newblock \emph{\bibinfo{journal}{J. Phys. Chem. B}}
  \textbf{\bibinfo{volume}{127}}, \bibinfo{pages}{6920--6927}
  (\bibinfo{year}{2023}).

\bibitem{larini_coarse-grained_2012}
\bibinfo{author}{Larini, L.} \& \bibinfo{author}{Shea, J.-E.}
\newblock \bibinfo{title}{Coarse-grained modeling of simple molecules at
  different resolutions in the absence of good sampling}.
\newblock \emph{\bibinfo{journal}{J. Phys. Chem. B}}
  \textbf{\bibinfo{volume}{116}}, \bibinfo{pages}{8337--8349}
  (\bibinfo{year}{2012}).

\bibitem{wang2021multi}
\bibinfo{author}{Wang, J.} \emph{et~al.}
\newblock \bibinfo{title}{Multi-body effects in a coarse-grained protein force
  field}.
\newblock \emph{\bibinfo{journal}{J. Chem. Phys.}}
  \textbf{\bibinfo{volume}{154}} (\bibinfo{year}{2021}).

\bibitem{ouyang_modelling_2015}
\bibinfo{author}{Ouyang, J.~F.} \& \bibinfo{author}{Bettens, R. P.~A.}
\newblock \bibinfo{title}{Modelling water: A lifetime enigma}.
\newblock \emph{\bibinfo{journal}{{CHIMIA}}} \textbf{\bibinfo{volume}{69}},
  \bibinfo{pages}{104--104} (\bibinfo{year}{2015}).

\bibitem{molinero_water_2009}
\bibinfo{author}{Molinero, V.} \& \bibinfo{author}{Moore, E.~B.}
\newblock \bibinfo{title}{Water modeled as an intermediate element between
  carbon and silicon}.
\newblock \emph{\bibinfo{journal}{J. Phys. Chem. B}}
  \textbf{\bibinfo{volume}{113}}, \bibinfo{pages}{4008--4016}
  (\bibinfo{year}{2009}).

\bibitem{noid2008mscg1}
\bibinfo{author}{Noid, W.~G.} \emph{et~al.}
\newblock \bibinfo{title}{{The multiscale coarse-graining method. I. A rigorous
  bridge between atomistic and coarse-grained models}}.
\newblock \emph{\bibinfo{journal}{J. Chem. Phys.}}
  \textbf{\bibinfo{volume}{128}}, \bibinfo{pages}{244114}
  (\bibinfo{year}{2008}).

\bibitem{larini2010multiscale}
\bibinfo{author}{Larini, L.}, \bibinfo{author}{Lu, L.} \&
  \bibinfo{author}{Voth, G.~A.}
\newblock \bibinfo{title}{The multiscale coarse-graining method. {VI}.
  implementation of three-body coarse-grained potentials}.
\newblock \emph{\bibinfo{journal}{J. Chem. Phys.}}
  \textbf{\bibinfo{volume}{132}} (\bibinfo{year}{2010}).

\bibitem{das_multiscale_2012}
\bibinfo{author}{Das, A.} \& \bibinfo{author}{Andersen, H.~C.}
\newblock \bibinfo{title}{The multiscale coarse-graining method. {IX}. a
  general method for construction of three body coarse-grained force fields}.
\newblock \emph{\bibinfo{journal}{J. Chem. Phys.}}
  \textbf{\bibinfo{volume}{136}}, \bibinfo{pages}{194114}
  (\bibinfo{year}{2012}).

\bibitem{schutt2021equivariant}
\bibinfo{author}{Sch{\"u}tt, K.}, \bibinfo{author}{Unke, O.} \&
  \bibinfo{author}{Gastegger, M.}
\newblock \bibinfo{title}{Equivariant message passing for the prediction of
  tensorial properties and molecular spectra}.
\newblock In \emph{\bibinfo{booktitle}{Int. Conf. Mach. Learn.}},
  \bibinfo{pages}{9377--9388} (\bibinfo{organization}{PMLR},
  \bibinfo{year}{2021}).

\bibitem{schutt2023schnetpack}
\bibinfo{author}{Sch{\"u}tt, K.~T.}, \bibinfo{author}{Hessmann, S.~S.},
  \bibinfo{author}{Gebauer, N.~W.}, \bibinfo{author}{Lederer, J.} \&
  \bibinfo{author}{Gastegger, M.}
\newblock \bibinfo{title}{Schnetpack 2.0: A neural network toolbox for
  atomistic machine learning}.
\newblock \emph{\bibinfo{journal}{J. Chem. Phys.}}
  \textbf{\bibinfo{volume}{158}} (\bibinfo{year}{2023}).

\bibitem{ercolessi_interatomic_1994}
\bibinfo{author}{Ercolessi, F.} \& \bibinfo{author}{Adams, J.~B.}
\newblock \bibinfo{title}{Interatomic potentials from first-principles
  calculations: The force-matching method}.
\newblock \emph{\bibinfo{journal}{{EPL}}} \textbf{\bibinfo{volume}{26}},
  \bibinfo{pages}{583} (\bibinfo{year}{1994}).

\bibitem{izvekov_multiscale_2005}
\bibinfo{author}{Izvekov, S.} \& \bibinfo{author}{Voth, G.~A.}
\newblock \bibinfo{title}{A multiscale coarse-graining method for biomolecular
  systems}.
\newblock \emph{\bibinfo{journal}{J. Phys. Chem. B}}
  \textbf{\bibinfo{volume}{109}}, \bibinfo{pages}{2469--2473}
  (\bibinfo{year}{2005}).

\bibitem{wang_machine_2019}
\bibinfo{author}{Wang, J.} \emph{et~al.}
\newblock \bibinfo{title}{Machine learning of coarse-grained molecular dynamics
  force fields}.
\newblock \emph{\bibinfo{journal}{{ACS} Cent. Sci.}}
  \textbf{\bibinfo{volume}{5}}, \bibinfo{pages}{755--767}
  (\bibinfo{year}{2019}).

\bibitem{mcquarrie76a}
\bibinfo{author}{McQuarrie, D.~A.}
\newblock \emph{\bibinfo{title}{Statistical Mechanics}}.
\newblock Harper's chemistry series (\bibinfo{publisher}{Harper Collins},
  \bibinfo{address}{New York}, \bibinfo{year}{1976}).

\bibitem{stock_unraveling_2017}
\bibinfo{author}{Stock, P.} \emph{et~al.}
\newblock \bibinfo{title}{Unraveling hydrophobic interactions at the molecular
  scale using force spectroscopy and molecular dynamics simulations}.
\newblock \emph{\bibinfo{journal}{{ACS} Nano}} \textbf{\bibinfo{volume}{11}},
  \bibinfo{pages}{2586--2597} (\bibinfo{year}{2017}).

\bibitem{monroe_decoding_2019}
\bibinfo{author}{Monroe, J.~I.} \& \bibinfo{author}{Shell, M.~S.}
\newblock \bibinfo{title}{Decoding signatures of structure, bulk
  thermodynamics, and solvation in three-body angle distributions of rigid
  water models}.
\newblock \emph{\bibinfo{journal}{J. Chem. Phys.}}
  \textbf{\bibinfo{volume}{151}}, \bibinfo{pages}{094501}
  (\bibinfo{year}{2019}).

\bibitem{scherer2018understanding}
\bibinfo{author}{Scherer, C.} \& \bibinfo{author}{Andrienko, D.}
\newblock \bibinfo{title}{Understanding three-body contributions to
  coarse-grained force fields}.
\newblock \emph{\bibinfo{journal}{Phys. Chem. Chem. Phys.}}
  \textbf{\bibinfo{volume}{20}}, \bibinfo{pages}{22387--22394}
  (\bibinfo{year}{2018}).

\bibitem{perez2013identification}
\bibinfo{author}{Pérez-Hernández, G.}, \bibinfo{author}{Paul, F.},
  \bibinfo{author}{Giorgino, T.}, \bibinfo{author}{De~Fabritiis, G.} \&
  \bibinfo{author}{Noé, F.}
\newblock \bibinfo{title}{Identification of slow molecular order parameters for
  markov model construction}.
\newblock \emph{\bibinfo{journal}{J. Chem. Phys.}}
  \textbf{\bibinfo{volume}{139}}, \bibinfo{pages}{015102}
  (\bibinfo{year}{2013}).

\bibitem{horng_characterization_2002}
\bibinfo{author}{Horng, J.-C.}, \bibinfo{author}{Moroz, V.},
  \bibinfo{author}{Rigotti, D.~J.}, \bibinfo{author}{Fairman, R.} \&
  \bibinfo{author}{Raleigh, D.~P.}
\newblock \bibinfo{title}{Characterization of large peptide fragments derived
  from the {N}-terminal domain of the ribosomal protein {L9}: Definition of the
  minimum folding motif and characterization of local electrostatic
  interactions}.
\newblock \emph{\bibinfo{journal}{Biochemistry}} \textbf{\bibinfo{volume}{41}},
  \bibinfo{pages}{13360--13369} (\bibinfo{year}{2002}).

\bibitem{anil_fine_2005}
\bibinfo{author}{Anil, B.}, \bibinfo{author}{Sato, S.}, \bibinfo{author}{Cho,
  J.-H.} \& \bibinfo{author}{Raleigh, D.~P.}
\newblock \bibinfo{title}{Fine structure analysis of a protein folding
  transition state; distinguishing between hydrophobic stabilization and
  specific packing}.
\newblock \emph{\bibinfo{journal}{J. Mol. Biol.}}
  \textbf{\bibinfo{volume}{354}}, \bibinfo{pages}{693--705}
  (\bibinfo{year}{2005}).

\bibitem{pymol}
\bibinfo{author}{Schr{\"o}dinger, L.} \& \bibinfo{author}{DeLano, W.}
\newblock \bibinfo{title}{{PyMOL} v2.5.0}.
\newblock \urlprefix\url{http://www.pymol.org/pymol}.

\bibitem{snow_kinetic_2006}
\bibinfo{author}{Snow, C.~D.}, \bibinfo{author}{Rhee, Y.~M.} \&
  \bibinfo{author}{Pande, V.~S.}
\newblock \bibinfo{title}{Kinetic definition of protein folding transition
  state ensembles and reaction coordinates}.
\newblock \emph{\bibinfo{journal}{Biophys. J.}} \textbf{\bibinfo{volume}{91}},
  \bibinfo{pages}{14--24} (\bibinfo{year}{2006}).

\bibitem{voelz2010molecular}
\bibinfo{author}{Voelz, V.~A.}, \bibinfo{author}{Bowman, G.~R.},
  \bibinfo{author}{Beauchamp, K.} \& \bibinfo{author}{Pande, V.~S.}
\newblock \bibinfo{title}{Molecular simulation of ab initio protein folding for
  a millisecond folder {NTL9} (1-39)}.
\newblock \emph{\bibinfo{journal}{J. Am. Chem. Soc.}}
  \textbf{\bibinfo{volume}{132}}, \bibinfo{pages}{1526--1528}
  (\bibinfo{year}{2010}).

\bibitem{lindorff-larsen_how_2011}
\bibinfo{author}{Lindorff-Larsen, K.}, \bibinfo{author}{Piana, S.},
  \bibinfo{author}{Dror, R.~O.} \& \bibinfo{author}{Shaw, D.~E.}
\newblock \bibinfo{title}{How fast-folding proteins fold}.
\newblock \emph{\bibinfo{journal}{Science}} \textbf{\bibinfo{volume}{334}},
  \bibinfo{pages}{517--520} (\bibinfo{year}{2011}).

\bibitem{sato2017}
\bibinfo{author}{Sato, S.}, \bibinfo{author}{Cho, J.-H.},
  \bibinfo{author}{Peran, I.}, \bibinfo{author}{Soydaner-Azeloglu, R.~G.} \&
  \bibinfo{author}{Raleigh, D.~P.}
\newblock \bibinfo{title}{The {N}-terminal domain of ribosomal protein {L9}
  folds via a diffuse and delocalized transition state}.
\newblock \emph{\bibinfo{journal}{Biophys. J.}} \textbf{\bibinfo{volume}{112}},
  \bibinfo{pages}{1797--1806} (\bibinfo{year}{2017}).

\bibitem{unke2021review}
\bibinfo{author}{Unke, O.~T.} \emph{et~al.}
\newblock \bibinfo{title}{Machine learning force fields}.
\newblock \emph{\bibinfo{journal}{Chem. Rev.}} \textbf{\bibinfo{volume}{121}},
  \bibinfo{pages}{10142--10186} (\bibinfo{year}{2021}).

\bibitem{lyubartsev_calculation_1995}
\bibinfo{author}{Lyubartsev, A.~P.} \& \bibinfo{author}{Laaksonen, A.}
\newblock \bibinfo{title}{Calculation of effective interaction potentials from
  radial distribution functions: A reverse monte carlo approach}.
\newblock \emph{\bibinfo{journal}{Phys. Rev. E}} \textbf{\bibinfo{volume}{52}},
  \bibinfo{pages}{3730--3737} (\bibinfo{year}{1995}).

\bibitem{lyubartsev_systematic_2009}
\bibinfo{author}{Lyubartsev, A.}, \bibinfo{author}{Mirzoev, A.},
  \bibinfo{author}{Chen, L.} \& \bibinfo{author}{Laaksonen, A.}
\newblock \bibinfo{title}{Systematic coarse-graining of molecular models by the
  newton inversion method}.
\newblock \emph{\bibinfo{journal}{Faraday Discuss.}}
  \textbf{\bibinfo{volume}{144}}, \bibinfo{pages}{43--56}
  (\bibinfo{year}{2009}).

\bibitem{ruhle2009versatile}
\bibinfo{author}{Ruhle, V.}, \bibinfo{author}{Junghans, C.},
  \bibinfo{author}{Lukyanov, A.}, \bibinfo{author}{Kremer, K.} \&
  \bibinfo{author}{Andrienko, D.}
\newblock \bibinfo{title}{Versatile object-oriented toolkit for coarse-graining
  applications}.
\newblock \emph{\bibinfo{journal}{J. Chem. Theory Comput.}}
  \textbf{\bibinfo{volume}{5}}, \bibinfo{pages}{3211--3223}
  (\bibinfo{year}{2009}).

\bibitem{montavon2019layer}
\bibinfo{author}{Montavon, G.}, \bibinfo{author}{Binder, A.},
  \bibinfo{author}{Lapuschkin, S.}, \bibinfo{author}{Samek, W.} \&
  \bibinfo{author}{M{\"u}ller, K.-R.}
\newblock \bibinfo{title}{Layer-wise relevance propagation: An overview}.
\newblock In \bibinfo{editor}{Samek, W.}, \bibinfo{editor}{Montavon, G.},
  \bibinfo{editor}{Vedaldi, A.}, \bibinfo{editor}{Hansen, L.~K.} \&
  \bibinfo{editor}{M{\"u}ller, K.-R.} (eds.)
  \emph{\bibinfo{booktitle}{Explainable {AI}: Interpreting, Explaining and
  Visualizing Deep Learning}}, \bibinfo{pages}{193--209}
  (\bibinfo{publisher}{Springer International Publishing},
  \bibinfo{year}{2019}).

\bibitem{montavon2017explaining}
\bibinfo{author}{Montavon, G.}, \bibinfo{author}{Lapuschkin, S.},
  \bibinfo{author}{Binder, A.}, \bibinfo{author}{Samek, W.} \&
  \bibinfo{author}{M{\"u}ller, K.-R.}
\newblock \bibinfo{title}{Explaining nonlinear classification decisions with
  deep taylor decomposition}.
\newblock \emph{\bibinfo{journal}{Pattern recognition}}
  \textbf{\bibinfo{volume}{65}}, \bibinfo{pages}{211--222}
  (\bibinfo{year}{2017}).

\bibitem{montavon2018methods}
\bibinfo{author}{Montavon, G.}, \bibinfo{author}{Samek, W.} \&
  \bibinfo{author}{M{\"u}ller, K.-R.}
\newblock \bibinfo{title}{Methods for interpreting and understanding deep
  neural networks}.
\newblock \emph{\bibinfo{journal}{Digit. Signal Process.}}
  \textbf{\bibinfo{volume}{73}}, \bibinfo{pages}{1--15} (\bibinfo{year}{2018}).

\bibitem{lapuschkin2019unmasking}
\bibinfo{author}{Lapuschkin, S.} \emph{et~al.}
\newblock \bibinfo{title}{Unmasking clever hans predictors and assessing what
  machines really learn}.
\newblock \emph{\bibinfo{journal}{Nat. Commun.}} \textbf{\bibinfo{volume}{10}},
  \bibinfo{pages}{1096} (\bibinfo{year}{2019}).

\bibitem{letzgus2024xpertai}
\bibinfo{author}{Letzgus, S.}, \bibinfo{author}{M{\"u}ller, K.-R.} \&
  \bibinfo{author}{Montavon, G.}
\newblock \bibinfo{title}{{XpertAI}: uncovering model strategies for
  sub-manifolds} (\bibinfo{year}{2024}).
\newblock \bibinfo{note}{Preprint at \url{https://arxiv.org/abs/2403.07486}}.

\bibitem{spooner2021counterfactual}
\bibinfo{author}{Spooner, T.} \emph{et~al.}
\newblock \bibinfo{title}{Counterfactual explanations for arbitrary regression
  models} (\bibinfo{year}{2021}).
\newblock \bibinfo{note}{Preprint at \url{https://arxiv.org/abs/2106.15212}}.

\bibitem{wang_comparative_2009}
\bibinfo{author}{Wang, H.}, \bibinfo{author}{Junghans, C.} \&
  \bibinfo{author}{Kremer, K.}
\newblock \bibinfo{title}{Comparative atomistic and coarse-grained study of
  water: What do we lose by coarse-graining?}
\newblock \emph{\bibinfo{journal}{Eur. Phys. J. E}}
  \textbf{\bibinfo{volume}{28}}, \bibinfo{pages}{221--229}
  (\bibinfo{year}{2009}).

\bibitem{eberle2020building}
\bibinfo{author}{Eberle, O.} \emph{et~al.}
\newblock \bibinfo{title}{Building and interpreting deep similarity models}.
\newblock \emph{\bibinfo{journal}{IEEE Trans Pattern Anal Mach Intell}}
  \textbf{\bibinfo{volume}{44}}, \bibinfo{pages}{1149--1161}
  (\bibinfo{year}{2020}).

\bibitem{ying2019gnnexplainer}
\bibinfo{author}{Ying, Z.}, \bibinfo{author}{Bourgeois, D.},
  \bibinfo{author}{You, J.}, \bibinfo{author}{Zitnik, M.} \&
  \bibinfo{author}{Leskovec, J.}
\newblock \bibinfo{title}{{GNNExplainer}: Generating explanations for graph
  neural networks}.
\newblock In \emph{\bibinfo{booktitle}{Adv. Neural Inf. Process. Syst.s}},
  vol.~\bibinfo{volume}{32} (\bibinfo{publisher}{Curran Associates, Inc.},
  \bibinfo{year}{2019}).

\bibitem{xiong2022efficient}
\bibinfo{author}{Xiong, P.}, \bibinfo{author}{Schnake, T.},
  \bibinfo{author}{Montavon, G.}, \bibinfo{author}{M{\"u}ller, K.-R.} \&
  \bibinfo{author}{Nakajima, S.}
\newblock \bibinfo{title}{Efficient computation of higher-order subgraph
  attribution via message passing}.
\newblock In \emph{\bibinfo{booktitle}{Int. Conf. Mach. Learn.}},
  \bibinfo{pages}{24478--24495} (\bibinfo{organization}{PMLR},
  \bibinfo{year}{2022}).

\bibitem{blucher2022preddiff}
\bibinfo{author}{Bl{\"u}cher, S.}, \bibinfo{author}{Vielhaben, J.} \&
  \bibinfo{author}{Strodthoff, N.}
\newblock \bibinfo{title}{{PredDiff}: Explanations and interactions from
  conditional expectations}.
\newblock \emph{\bibinfo{journal}{Artif. Intell.}}
  \textbf{\bibinfo{volume}{312}}, \bibinfo{pages}{103774}
  (\bibinfo{year}{2022}).

\bibitem{faber2021comparing}
\bibinfo{author}{Faber, L.}, \bibinfo{author}{K.~Moghaddam, A.} \&
  \bibinfo{author}{Wattenhofer, R.}
\newblock \bibinfo{title}{When comparing to ground truth is wrong: On
  evaluating gnn explanation methods}.
\newblock In \emph{\bibinfo{booktitle}{Proceedings of the 27th ACM SIGKDD
  conference on knowledge discovery \& data mining}}, \bibinfo{pages}{332--341}
  (\bibinfo{year}{2021}).

\bibitem{janizek2021explaining}
\bibinfo{author}{Janizek, J.~D.}, \bibinfo{author}{Sturmfels, P.} \&
  \bibinfo{author}{Lee, S.-I.}
\newblock \bibinfo{title}{Explaining explanations: Axiomatic feature
  interactions for deep networks}.
\newblock \emph{\bibinfo{journal}{J. Mach. Learn. Res.}}
  \textbf{\bibinfo{volume}{22}}, \bibinfo{pages}{1--54} (\bibinfo{year}{2021}).

\bibitem{lundberg2020local}
\bibinfo{author}{Lundberg, S.~M.} \emph{et~al.}
\newblock \bibinfo{title}{From local explanations to global understanding with
  explainable {AI} for trees}.
\newblock \emph{\bibinfo{journal}{Nat. Mach. Intell.}}
  \textbf{\bibinfo{volume}{2}}, \bibinfo{pages}{56--67} (\bibinfo{year}{2020}).

\bibitem{dhamdhere2019shapley}
\bibinfo{author}{Sundararajan, M.}, \bibinfo{author}{Dhamdhere, K.} \&
  \bibinfo{author}{Agarwal, A.}
\newblock \bibinfo{title}{The shapley taylor interaction index}.
\newblock In \emph{\bibinfo{booktitle}{Proceedings of the 37th International
  Conference on Machine Learning}}, \bibinfo{pages}{9259--9268}
  (\bibinfo{publisher}{{PMLR}}, \bibinfo{year}{2020}).

\end{thebibliography}

\newpage

\end{document}